\documentclass[conference]{IEEEtran}
\IEEEoverridecommandlockouts
\usepackage{cite}
\usepackage{amsmath,amssymb,amsfonts}
\usepackage{bm}
\usepackage{dsfont}
\usepackage{algorithmic}
\usepackage{graphicx}
\usepackage{textcomp}
\usepackage{xcolor}
\usepackage{orcidlink}
\usepackage{siunitx}
\usepackage{tikz}
\usetikzlibrary{calc}
\usepackage{pgfplots}

\makeatletter
\let\MYcaption\@makecaption
\makeatother

\usepackage[font=footnotesize]{subcaption}

\makeatletter
\let\@makecaption\MYcaption
\makeatother

\def\BibTeX{{\rm B\kern-.05em{\sc i\kern-.025em b}\kern-.08em
		T\kern-.1667em\lower.7ex\hbox{E}\kern-.125emX}}
\begin{document}
	
	\title{Anomaly Detection for Sensing Security  
		\thanks{This work was supported by the German Federal Ministry of Research, Technology and Space (BMFTR) project 6G-ANNA [grant agreement number 16KISK095].}%
	}
	
	\author{Stefan Roth\,\orcidlink{0000-0003-4408-3306} and Aydin Sezgin\,\orcidlink{0000-0003-3511-2662}%
		\thanks{S. Roth and A. Sezgin are with the Institute of Digital Communication Systems, Ruhr University Bochum, 44801 Bochum, Germany (e-mail: stefan.roth-k21@rub.de, aydin.sezgin@rub.de).}}
	
	\maketitle
	
	\begin{abstract}
		Various approaches in the field of physical layer security involve anomaly detection, such as physical layer authentication, sensing attacks, and anti-tampering solutions. Depending on the context in which these approaches are applied, anomaly detection needs to be computationally lightweight, resilient to changes in temperature and environment, and robust against phase noise. We adapt moving average filters, autoregression filters and Kalman filters to provide predictions of feature vectors that fulfill the above criteria. Different hypothesis test designs are employed that allow omnidirectional and unidirectional outlier detection. In a case study, a sensing attack is investigated that employs the described algorithms with various channel features based on commodity WiFi devices. Thereby, various combinations of algorithms and channel features show effectiveness for motion detection by an attacker. Countermeasures only utilizing transmit power randomization are shown insufficient to mitigate such attacks if the attacker has access to channel state information (CSI) measurements, suggesting that mitigation solutions might require frequency-variant randomization.
	\end{abstract}
	
	
	\section{Introduction}
	Physical layer security is a broad field, within which various attacks and security measures for confidentiality, authenticity, tamperproofing, and privacy against localization and sensing attacks have been proposed \cite{Roth2024}. Although the proposed measures vary significantly in scope and methodology, multiple of such measures applied by attackers and defenders depend on the detection of anomalies:
	\begin{itemize}
		\item In \emph{physical layer authentication}, defenders aim to detect anomalies in the form of packets coming from an adversary to distinguish between legitimate and adversarial packages \cite{Roth2024,9984839}.
		\item In \emph{sensing attacks} (also referred to as \emph{radio window attacks}), attackers aim to sense information from the environment. For example, if the attacker is interested in detecting motion of legitimate wireless devices or people in the surrounding of the devices, it needs to detect anomalies in the form of present motion \cite{Zhu2018EtTA,Staat2021IRShieldAC,10.1145/3395351.3399368}.
		\item In \emph{anti-tampering solutions}, defenders monitor tampering attacks against physical devices and need to detect anomalies that occur when a device has been opened or physically affected in another way \cite{9833631}. 
	\end{itemize}
	Within all of such attack and defense mechanisms, test designs are employed that distinguish between the regular operation and anomalies. Therefore, often an initialization phase is employed as a reference, in which parameters such as a feature prediction are calibrated, while other measures guarantee that no attack or motion occurs \cite{Zhu2018EtTA,Staat2021IRShieldAC,9833631}. However, material properties are sensitive to temperature changes and also the environment of the devices can change over time. Hence, mechanisms that can adapt to such changes are needed. Especially, static feature predictions are not suitable if there are slow changes in the environments. Instead, algorithms need to distinguish between slow changes created by variations in temperature or environment and rapid changes due to the described anomalies.
	
	The data features used for the anomaly detection within proposals for the above security mechanisms and attacks vary across the literature. Multiple works directly or indirectly employ measurements of the wireless channels as such features. For example, in \cite{Roth2024,9984839}, complex-valued channel state information (CSI) has been used to provide authenticity. As phase information is often incoherent between different measurements, an option is to estimate and correct the phases algorithmically \cite{9034987}. Alternatively, the metric used to evaluate the channel can be chosen such that only the amplitude of the channel or the received signal strength indicator is relevant \cite{9833631,Zhu2018EtTA,Staat2021IRShieldAC}. In the sensing attack from \cite{Zhu2018EtTA,Staat2021IRShieldAC}, the amplitudes of a few channel measurements are combined to calculate a short-time variance which is then used as feature for the anomaly detection. Other works employ hardware impairments as features. For instance, the inphase/quadrature imbalance and carrier frequency offset are employed for authentication in \cite{7036875} and \cite{6804410}, respectively. Similarly, sensor measurements are used as a feature in \cite{9595413,9984839}. In all of the above cases, the features are evaluated in mechanisms that involve tracking and a hypothesis test. Also, the parameters of the hypothesis test can be calculated from prior measurements.
	
	As discussed above, the reliance on a static feature prediction often has practical issues in real world implementations such as a low resilience to temperature changes. Moreover, where anomaly detection algorithms based on complex-valued CSI are employed in practice, algorithms need to be chosen that are robust against phase noise. Furthermore, depending on the security target, a lightweight design of the hypothesis test might be required. For instance, physical layer authentication schemes often need to operate significantly faster than cryptography approaches to protect against denial of service (DoS) attacks. As moving average filters, autoregression filters, and Kalman filters are lightweight algorithms frequently used for signal processing, we focus here on adapting these algorithms for resilience and robustness.
	
	\paragraph*{Contribution} The main contribution is as follows:
	\begin{itemize}
		\item We provide different mechanisms for anomaly detection in the presence of phase noise. Therefore, we start by predicting new feature values using moving average, autoregressive, and Kalman filters. After this, we feed the predictions into optimized hypothesis tests that detect the anomalies.
		\item We implement the proposed algorithms in a case study of a sensing attack in which an adversary aims to sense motion. Based on experimental measurements, different channel features are calculated based on the CSI and received signal strength indicator (RSSI) measurements, which are fed into the proposed algorithms to sense motion in real time. Numerical results showing the motion detection performance with all proposed algorithms and considered features are provided.
	\end{itemize}
	
	\paragraph*{Notation} Scalars, vectors, and matrices are indicated in italic, lower-case bold-face, and upper-case bold-face letters. $\mathcal{N}(\bm{\mu},\bm{V})$ is the real-valued or complex Gaussian distribution, while $\bm{U}([a,b])$ refers to the equal distribution.  $\bm{I}_K$ is the identity matrix of size $K$, while $j$ is the imaginary unit. $\circ$ denotes the Hadamard product. $\mathrm{diag}(\bm{A})$ is the vector containing all elements of the diagonal of $\bm{A}$.

	\section{System Model}
	
	We consider a system, within which vectors of data features $\bm{y}(t)$ are generated at time instances $t\in\{0,1,2,\dots\}$, among which anomalies should be detected. Feature vectors of normal measurements are referred to as $\bm{y}_0(t)$, while the data features of measurements containing anomalies are denoted as $\bm{y}_1(t)$ in the following. For the normal measurements, we assume that the data features $\bm{y}_0(t)$ are represented by an $K$-element long vector and follow a Gauss-Markov model. This assumption is accurate for wireless channels that follow a Rayleigh fading model with a Gauss-Markov distributed channel state. Within each time instant $t$, the measurement $\bm{y}_0(t)$ can be described as
	\begin{align}
		\bm{x}_0(t+1)&=a\bm{x}_0(t)+\bm{u}(t),\label{eq:system:state}\\
		\bm{y}_0(t)&=e^{j\phi(t)}\bm{x}_0(t)+\bm{v}(t),\label{eq:system:output}\\
		\bm{x}_0(0)&=\bm{x}_{0,\mathrm{init}}.
	\end{align}
	Thereby, $a$ determines how the previous feature values impact the follow-up measurements. In most systems, $a\lessapprox 1$ holds, so that the feature vector only has minor changes between consecutive feature measurements and the feature values are stable. The vector $\bm{u}(t)\sim\mathcal{N}(\bm{0},\bm{U})$ describes the new part of the features at each time instant, of which $\bm{U}$ is the covariance matrix. $\bm{x}_{0,\mathrm{init}}$ is the initial state at time $0$, which is randomly distributed following $\mathcal{N}\left(0,\bm{X}_{0,\mathrm{init}}\right)$ and finite. $\phi(t)\sim\mathcal{U}([0,2\pi))$ indicates random phase offsets potential occurring within the feature measurements\footnote{While some practical systems additionally contain a phase drift within CSI data ranging over different subcarriers \cite{9034987}, the phase drift is assumed to be negligible within the scope of this work.}.
	
	For anomaly packets, we assume that no tracking is possible, as anomalies occur rarely or have a distinct pattern. Hence, no prior knowledge can be assumed available and the variance of these packets can be significantly larger than the variance of regular packets. A suitable model is
	\begin{align}
		\bm{y}_1(t)&~\sim\mathcal{N}(\gamma\bm{y}_0(t),\bm{Y}),
	\end{align}
	where $\gamma\in\{0,1\}$ indicates if these random variables are centered around the origin or the regular feature value and $\bm{Y}$ is the covariance matrix. Note that, depending on the data type, $\mathcal{N}$ can be either the real-valued Gaussian distribution or complex-valued Gaussian distribution.
	
	When evaluating the feature vector $\bm{y}(t)$, we assume a prediction of $\bm{y}_0(t)$ to be available referred to as $\bm{\hat{y}}_0(t)$, which we will discuss later. As in principle multiple test designs can possibly be used, we focus on a general metric $h(\bm{y}(t)|\bm{\hat{y}}_0(t))$. If this metric is low, that is, if $\bm{y}(t)$ is similar to the prediction, it is assumed that there is no anomaly. Hence, we get a hypothesis test of
	\begin{align}
		H(t)=\begin{cases}H_0 & h(\bm{y}(t)|\bm{\hat{y}}_0(t)) \leq \eta,\\H_1 & h(\bm{y}(t)|\bm{\hat{y}}_0(t)) > \eta.\end{cases}\label{eq:test}
	\end{align}
	Thereby, $\eta$ is the threshold value.

	\section{Test Optimization}
	While the hypothesis test in \eqref{eq:test} can in principle be arbitrarily designed, we focus on test designs based on moving average prediction, autoregressive prediction, and Kalman filtering. Therefore, we first predict the new feature vector $\bm{y}(t)$ as $\bm{\hat{y}}_0(t)$ and describe the uncertainty of the estimates via the three methods in Subsection~\ref{sec:featurePrediction}. Based on this, in Subsection~\ref{sec:h}, we discuss possible designs of the functions $h(\bm{y}(t)|\bm{\hat{y}}_0(t))$ based on omnidirectional and unidirectional anomaly detection methods. Finally, we provide a few comments on the choice of $\eta$ in Subsection~\ref{sec:eta}.

	\subsection{Feature Prediction}\label{sec:featurePrediction}
	
	\subsubsection{Moving Average Prediction}\label{sec:MA}
	
	Within moving average predictions, the last $N$ feature measurements $\bm{y}(t-n)$ are used, $n\in\{1,\dots,N\}$. If all of these measurements have an equal weight and no phase noise occurs, the predicted feature vector can be formulated as
	\begin{align}
		\bm{\hat{y}}_0(t)=\frac{1}{N}\sum_{n=1}^N\bm{y}(t-n).
	\end{align}
	While a smaller value of $N$ better fits the dynamics in \eqref{eq:system:state}, a larger value of $N$ is used to ensure robustness against measurement noise.
	
	The handling of anomalies depends on the context, in which the anomaly detection is implemented. Some applications might enable a self-destruction mode, in which the device is de-activated until the issue is investigated manually \cite{9833631}. In other applications, the recognition of multiple anomalies is required, which, however, should not influence further measurements. Here, the test performance can further be improved by removing the terms related to the measurements from time instances, at which anomalies have been detected. However, a full removal could lead to the system becoming locked in a state, in which an increased background noise is continuously detected as anomaly, without any possibility of leaving this state again. Hence, the introduction of weights can be a useful choice, i.e.,
	\begin{align}
		\bm{\hat{y}}_0(t)&=\frac{\sum_{n=1}^N \alpha(t-n)\bm{y}(t-n)}{\sum_{n=1}^N \alpha(t-n)},
	\end{align}
	in which
	\begin{align}
		\alpha(t)=\begin{cases}
			\tilde{\alpha}_0 & H(t)=H_0,\\\tilde{\alpha}_1 & H(t)=H_1,
		\end{cases}\label{eq:alpha_n}
	\end{align}
	and $\tilde{\alpha}_0\gg \tilde{\alpha}_1$\footnote{An alternative the the method suggested here is to remove a certain percentage of the last $N$ feature measurements that have most likely been related to anomalies. In this case, it is not necessary to apply different weights as a full removal from the prediction is possible.}.
	
	If phase noise is present, the alignment of the different vectors is affected, which leads to suboptimal results. This could be handled by selecting one of the feature vectors as a reference for the phase and applying a correction term. When assuming that the reference feature vector is the one recorded at time $n'$, the term $\bm{\hat{y}}_0(t)$ can be reformulated as
	\begin{align}
		\bm{\hat{y}}_0(t)&=\frac{\sum_{n=1}^N \alpha(t-n)\frac{\bm{y}(t-n)^H\bm{y}(t-n')}{|\bm{y}(t-n)^H\bm{y}(t-n')|}\bm{y}(t-n)}{\sum_{n=1}^N \alpha(t-n)}.\label{eq:y_MR_3}
	\end{align}
	
	Similarly, the covariance matrix of the measurements $\bm{y}_0(t)$ can be obtained as
	\begin{align}
		\bm{\hat{Y}}_0(t)&=\frac{\sum_{n=1}^N \alpha(t-n)\bm{y}(t-n)\bm{y}(t-n)^H}{\sum_{n=1}^N \alpha(t-n)}-\bm{\hat{y}}_0(t)\bm{\hat{y}}_0(t)^H.\label{eq:Y_hat_0_MR}
	\end{align}
	
	\subsubsection{Autoregressive Prediction}\label{sec:AR}
	In autoregressive predictions, a prediction is iteratively updated with each new feature measurement $\bm{y}(t)$. For the first feature measurement, the prediction vector and correlation matrix can be estimated as $\bm{\hat{y}}_0(0)=\bm{0}_K$ and $\bm{\hat{\underline{Y}}}_0(0)=\bm{X}_{0,\mathrm{init}}+\bm{V}$, respectively. Based on a weight $\alpha$ that describes the relative importance between the latest measurement and the long-term trend, the predicted feature value can be iteratively updated as
	\begin{align}
		\bm{\hat{y}}_0(t)=\alpha\bm{y}(t-1)+(1-\alpha)\bm{\hat{y}}_0(t-1).\label{eq:y_hat_AR}
	\end{align}
	Depending on how anomalies are handled in the considered context, it can also here be useful to introduce weights similar to \eqref{eq:alpha_n} to avoid locked states. In the case that different weights are used depending on if anomalies have been detected, \eqref{eq:y_hat_AR} can be reformulated as
	\begin{align}
		\bm{\hat{y}}_0(t)=\alpha(t-1)\bm{y}(t-1)+(1-\alpha(t-1))\bm{\hat{y}}_0(t-1).\label{eq:y_hat_AR_2}
	\end{align}
	If phase noise is present, the feature vector $\bm{y}(t-1)$ has to be aligned with the previously available prediction $\bm{\hat{y}}_0(t-1)$. This can be achieved by reformulating \eqref{eq:y_hat_AR_2} to
	\begin{align}
		\bm{\hat{y}}_0(t)&=\alpha(t-1)\frac{\bm{y}(t-1)^H\bm{\hat{y}}_0(t-1)}{|\bm{y}(t-1)^H\bm{\hat{y}}_0(t-1)|}\bm{y}(t-1)\nonumber\\&\hspace{0.5cm}+(1-\alpha(t-1))\bm{\hat{y}}_0(t-1).\label{eq:y_hat_AR_3}
	\end{align}
	
	The correlation matrix $\bm{\hat{\underline{Y}}}_0(t+1)$ and covariance matrix $\bm{\hat{Y}}_0(t)$ that describe the uncertainity of the estimates can be approximated iteratively as
	\begin{align}
		\bm{\hat{\underline{Y}}}_0(t+1)&=\alpha(t)\bm{y}(t)\bm{y}(t)^H+(1-\alpha(t))\bm{\hat{\underline{Y}}}_0(t),\label{eq:A}\\
		\bm{\hat{Y}}_0(t)&=\bm{\hat{\underline{Y}}}_0(t)-\bm{\hat{y}}_0(t)\bm{\hat{y}}_0(t)^H.\label{eq:Y_hat_0_AR}
	\end{align}
	
	\subsubsection{Kalman Filter Prediction}\label{sec:Kalman}
	The Kalman filter \cite{Kalman1960} allows the iterative prediction of the feature values similar to the autoregressive filter discussed before. While the Kalman filter leads to potentially more accurate predictions, it comes at the cost that estimates of more model parameters need to be known than with the previous two methods. To initialize this filter, the feature value can be assumed at the beginning as $\bm{\hat{y}}_0(0)=\bm{0}_K$, while the covariance matrix of the feature state is $\bm{\hat{X}}_0(0)=\bm{X}_{0,\mathrm{init}}$.
	
	Based on each newly arriving measurement, the estimation can be improved as
	\begin{align}
		\bm{K}(t)&=\bm{\hat{X}}_0(t)\left(\bm{\hat{X}}_0(t)+\bm{\hat{V}}\right)^{-1},\label{eq:Kalman_K}\\
		\bm{\hat{y}}_0(t|t)&=\bm{\hat{y}}_0(t)+\bm{K}(t)\left(\bm{y}(t)-\bm{\hat{y}}_0(t)\right),\label{eq:Kalman_y_0_t_t}\\
		\bm{\hat{X}}_0(t|t)&=\left(\bm{I}_K-\bm{K}(t)\right)\bm{\hat{X}}_0(t).\label{eq:Kalman_X_0_t_t}
	\end{align}
	Thereby, $\bm{\hat{V}}$ is an estimation of $\bm{V}$. When the weights in \eqref{eq:alpha_n} are used in the handling of the anomalies, \eqref{eq:Kalman_K} changes to
	\begin{align}
		\bm{K}(t)&=\bm{\hat{X}}_0(t)\left(\bm{\hat{X}}_0(t)+\frac{\tilde{\alpha}_0}{\alpha(t)}\bm{\hat{V}}\right)^{-1}.
	\end{align}
	If phase noise is present, \eqref{eq:Kalman_y_0_t_t} becomes
	\begin{align}
		\bm{\hat{y}}_0(t|t)&=\bm{\hat{y}}_0(t)+\bm{K}(t)\left(\frac{\bm{y}(t)^H\bm{\hat{y}}_0(t)}{|\bm{y}(t)^H\bm{\hat{y}}_0(t)|}\bm{y}(t)-\bm{\hat{y}}_0(t)\right).
	\end{align}
	
	The improved estimates can then be used to predict follow-up feature state values as
	\begin{align}
		\bm{\hat{y}}_0(t)&=\hat{a}(t) \bm{\hat{y}}_0(t-1|t-1),\label{eq:Kalman:y_0_t}\\
		\bm{\hat{X}}_0(t)&=\hat{a}(t)^2 \bm{\hat{X}}_0(t-1|t-1)+\bm{\hat{U}}(t),
	\end{align}
	of which $\bm{\hat{U}}(t)$ and $\hat{a}(t)$ are the estimates of $\bm{U}$ and $a$ used at time instance $t$. 
	
	The Kalman filter requires estimates of the parameters $\hat{a}(t)$, $\bm{\hat{U}}(t)$ and $\bm{\hat{V}}$. If $\bm{\hat{U}}(t)$ is not known, an expectation maximization (EM) algorithm can be used to estimate the parameter $\bm{\hat{U}}(t)$ alternating with the calculation of the prediction. While different implementations of the EM algorithm are possible, we focus here on one of such implementations. Similar to \eqref{eq:Kalman_y_0_t_t} and \eqref{eq:Kalman:y_0_t}, the correlation matrix of the measurements can be estimated iteratively as
	\begin{align}
		\bm{\hat{\underline{Y}}}_0(t|t)&=\bm{\hat{\underline{Y}}}_0(t)+\bm{K}(t)\left(\bm{y}(t)\bm{y}(t)^H-\bm{\hat{\underline{Y}}}_0(t)\right),\label{eq:Kalman_Z_0_t_t}\\
		\bm{\hat{\underline{Y}}}_0(t)&=\hat{a}(t)^2 \bm{\hat{\underline{Y}}}_0(t-1|t-1)+(1-\hat{a}(t)^2)\bm{\hat{V}}+\bm{\hat{U}}(t).\label{eq:Kalman:Z_0_t}
	\end{align}
	From this, the covariance matrix $\bm{\hat{Y}}_0(t)$ can be calculated as described in \eqref{eq:Y_hat_0_AR}. As initialization for $\bm{\hat{\underline{Y}}}_0(0)$, an estimate for $\bm{X}_{0,\mathrm{init}}+\bm{V}$ needs to be selected. Under the assumption that the measurement noise $\bm{V}$ is negligible, the correlation matrix of $\bm{x}_0(t)$ can be assumed equal to $\bm{\hat{\underline{Y}}}_0(t)$. From the static case of $\bm{\underline{Y}}_0=\sum_{i=0}^\infty a^{2i}\bm{U}=(1-\hat{a}(t)^2)^{-1}\bm{U}$, we can estimate the covariance matrix of the new part of the feature as
	\begin{align}
		\bm{\hat{U}}(t)&=(1-\hat{a}(t)^2)\bm{\hat{\underline{Y}}}_0(t).
	\end{align}
	The parameters $\bm{\hat{V}}$ can be selected arbitrarily as values fitting well to the data. The parameter $\hat{a}(t)$ can be chosen as $1-\alpha(t)$ and fulfills $\hat{a}(t)\lessapprox1$. When $\bm{X}_{0,\mathrm{init}}$, $\bm{\hat{V}}$, $\bm{\hat{U}}(t)$, and $\bm{y}(t)\bm{y}(t)^H$ are approximated as diagonal, parts of the calculations simplify including the inversion in the calculation of $\bm{K}(t)$.

	\subsection{The Hypothesis Test Design}\label{sec:h}
	
	\subsubsection{Omnidirectional Anomalies}
	
	Based on any of the above formulations of the estimator $\bm{\hat{y}}_0(t)$, the function $h(\bm{y}(t)|\bm{\hat{y}}_0(t))$ can be formulated. When omnidirectional anomalies, i.e., feature vectors with larger and smaller amplitudes than $\bm{\hat{y}}_0(t)$ should be found, $h(\bm{y}(t)|\bm{\hat{y}}_0(t))$ can be formulated as \cite{9984839,Roth2024}
	\begin{align}
		h(\bm{y}(t)|\bm{\hat{y}}_0(t))&=\frac{1}{K}\Big(\left(\bm{y}(t)-\bm{\hat{y}}_0(t)\right)^H\bm{\Gamma}(t)\left(\bm{y}(t)-\bm{\hat{y}}_0(t)\right)\Big),\label{eq:h_1}
	\end{align}
	in which $\bm{\Gamma}(t)$ is a positive semi-definite Hermitian matrix. This formulation is well-suited for the case that no phase noise is present or only the real-valued amplitudes of the channel are investigated. However, if phase noise occurs, the measurements of $\bm{y}(t)$ can be misaligned, leading to falsified results in the hypothesis test. For the case of a phase offset, the phase of the new feature measurement $\bm{y}(t)$ and the prediction $\bm{\hat{y}}_0(t)$ need to be aligned. By doing so, we get
	\begin{align}
		h(\bm{y}(t)|\bm{\hat{y}}_0(t))
		&=\frac{1}{K}\Big(\bm{y}(t)^H\bm{\Gamma}(t)\bm{y}(t)-2\left|\bm{y}(t)^H\bm{\Gamma}(t)\bm{\hat{y}}_0(t)\right|\nonumber\\&\hphantom{=}+\bm{\hat{y}}_0(t)^H\bm{\Gamma}(t)\bm{\hat{y}}_0(t)\Big).\label{eq:h_2}
	\end{align}

	The matrix $\bm{\Gamma}(t)$ and the threshold value $\eta$ connect the probabilities of \eqref{eq:test} for ordinary and anomaly $\bm{y}(t)$ \cite{9984839}. Depending on which of these probabilities should be constrained, a different method of calculating $\bm{\Gamma}(t)$ should be selected.
	For normal packets without phase noise, the difference in \eqref{eq:h_1} has an expectation near zero, but a non-zero variance. Hence, in the case that a low detection is required for these packets, a suitable choice of $\bm{\Gamma}(t)$ is the inverse of the covariance matrix $\bm{\hat{Y}}_0(t)$, i.e., $\bm{\Gamma}(t)=\left(\bm{\hat{Y}}_0(t)\right)^{-1}$. If $K$ is large, the inversion can become numerically complex. As an alternative, this expression can be approximated as
	\begin{align}
		\bm{\Gamma}(t)&=\left(\bm{\hat{Y}}_0(t)\circ\bm{I}_K\right)^{-1}.\label{eq:Gamma_MR}
	\end{align}
	Within this expression, only the values of the diagonal of $\bm{\hat{Y}}_0(t)$ in \eqref{eq:Y_hat_0_MR} or \eqref{eq:y_hat_AR_3} need to be computed and can be inverted individually. If phase offsets are present, these cancel as both calculations of $\bm{\hat{Y}}_0(t)$ contain feature vectors only as a product with their own Hermitian. 
	
	\subsubsection{Unidirectional Anomalies}
	
	Parts of the existing sensing attacks use short-term channel variations as feature and target to detect only unidirectional anomalies larger than the prediction \cite{Zhu2018EtTA}. If $\bm{y}(t)$ is positive, $\bm{\Gamma}(t)$ is diagonal, and only positive anomalies need to be considered, $h(\bm{y}(t)|\bm{\hat{y}}_0(t))$ can alternatively be formulated as
	\begin{align}
		h(\bm{y}(t)|\bm{\hat{y}}_0(t))&=\frac{1}{K}\mathrm{diag}\left(\bm{\Gamma}(t)^{0.5}\right)^T\left(\bm{y}(t)-\bm{\hat{y}}_0(t)\right).\label{eq:h_3}
	\end{align}
	If phase noise is present, \eqref{eq:h_3} can be reformulated as
	\begin{align}
		h(\bm{y}(t)|\bm{\hat{y}}_0(t))&\nonumber\\&\hspace{-1.7cm}=\frac{1}{K}\mathrm{diag}\left(\bm{\Gamma}(t)^{0.5}\right)^T\left(\frac{\bm{y}(t)^H\bm{\hat{y}}_0(t)}{|\bm{y}(t)^H\bm{\hat{y}}_0(t)|}\bm{y}(t)-\bm{\hat{y}}_0(t)\right).\label{eq:h_4}
	\end{align}
	
	\subsection{The Threshold Value Choice} \label{sec:eta}
	With the above choices of $\bm{\Gamma}(t)$, the functions $h(\bm{y}(t)|\bm{\hat{y}}_0(t))$ equal for non-anomaly packets approximately the sum of (squared) normalized Gaussian variables divided by the number of Gaussian variables. From the $68-95-99.7$ rule, we obtain for the omnidirectional anomaly detection and $K=1$ that the value $\eta=1$ leads approximately to the filtering of $68.27\%$ of the normal packets, $\eta=4$ to the filtering of approximately $95.45\%$ and $\eta=9$ to the filtering of approximately $99.73\%$. For the unidirectional anomaly detection, approximately  $84.14\%$ ($\eta=1$), $97.73\%$ ($\eta=2$), or $99.87\%$ ($\eta=3$) of the normal packets are filtered. However, these are only rough approximates, as, e.g., the different values of the feature vectors can be correlated, estimation errors can be present, and deviations from the Gaussian model occur.
	
	To select a value of $\eta$ for a specific test design and feature type, initial measurements can be used from times at which no motion is present. For instance, the threshold values can be set to $\eta=\infty$ and the algorithm can be run while it is guaranteed with external measures that no anomalies occur. Once all parameters of the algorithm have converged, values of $h(\bm{y}(t)|\bm{\hat{y}}_0(t))$ can be recorded to find a value that is exceeded only in a small fraction of all measurements, which can then be used with a small tolerance margin as the threshold value. While this step also means that some initialization is required, the choice of $\eta$ has a significantly higher long-term stability and is more easy to be transferred to other environments than the feature value $\bm{y}_0(t)$.

	\section{Case Study: Sensing Attacks}
	Within sensing attacks, adversaries aim to use wireless signals to sense the environment to obtain privacy-sensitive information, such as on the presence of human motion \cite{Staat2021IRShieldAC}. Such attacks can be done actively, i.e., based on signals transmitted by the adversary, or passively, i.e., by overhearing signals broadcast by legitimate devices already present in the environment. In this case study, we implement a passive sensing attack based on the proposed algorithms on two TP-Link N750 wireless routers running the OpenWrt operation system and the Atheros CSI Tool \cite{8423070}. Each router is equipped with three antennas. We use the first WiFi router in a room to represent an arbitrary commodity WiFi device located in the protected area, which continuously transmits data packets in the \SI{5}{GHz} band. The second WiFi router is located outside of the room, which is where an adversary would have physical access, and used to measure the CSI and RSSI data from the wireless packets and to forward them to a PC running Matlab. The PC bundles the measurements from a time interval of \SI{0.2}{s}. Each bundle alone ($B=1$) and the combination of $B=10$ bundles together are used to calculate feature vectors from the data. The feature values considered in this work are as follows.
	\begin{itemize}
		\item The frequency-average standard deviation of the amplitude of the CSI value at each subcarrier, similar to \cite{Zhu2018EtTA}.
		\item The frequency-average variance of the amplitude of the CSI value at each subcarrier.
		\item The vector containing the standard deviation of the amplitude of the CSI values at each subcarrier.
		\item The vector containing the variance of the amplitude of the CSI values at each subcarrier.
		\item The vector containing the time-average amplitude of the CSI values.
		\item The scalar containing the standard deviation of the RSSI values.
		\item The scalar containing the variance of the RSSI values.
		\item The scalar containing the time-average RSSI values.
	\end{itemize}
	Directly after the recording of each bundle of measurement and the calculation of the feature vectors, the PC is used to evaluate the features at the PC through the different algorithms proposed above. As a short moment of time is needed for the computation, the data packets received during this time are neglected.
	
	Within all algorithms, the parameters $\tilde{\alpha}_0=0.02$ and $\tilde{\alpha}_1=0.001$ are used. The moving average prediction algorithm uses the last $N=200$ feature measurements to create its prediction. Within the Kalman filter, the parameter measurement noise estimate are chosen as $\bm{\hat{V}}=2\bm{I}_K$. All feature vectors contain either $K=1$ or $K=504$ elements. 
	\begin{figure}
		\centering
		\definecolor{mycolor1}{rgb}{0.97223,0.58172,0.25408}%
\definecolor{mycolor2}{rgb}{0.79638,0.27798,0.47132}%
\definecolor{mycolor3}{rgb}{0.49343,0.01152,0.65803}%
\definecolor{mycolor4}{rgb}{0.05038,0.02980,0.52797}%
\begin{tikzpicture}
    \draw[thick,fill=gray!10!white] (0,0)--(0,0.42)--(0.6,0.42)--(0.6,0)--(0,0);
    \draw[thick,fill=gray!10!white] (0.6,0)--(0.6,0.32)--(1.4,0.32)--(1.4,0)--(0.6,0);
    \draw[thick,fill=gray!10!white] (1.4,0)--(1.4,0.32)--(2.2,0.32)--(2.2,0)--(1.4,0);
    \draw[thick,fill=gray!10!white] (0,3.32)--(0,1.92)--(0.44,1.92)--(0.44,3.32)--(0,3.32);
    \draw[thick,fill=gray!10!white] (0,-3.57)--(0,-1.57)--(0.44,-1.57)--(0.44,-3.57)--(0,-3.57);

    \draw[thick,fill=gray!10!white] (1.01,-0.13)--(1.61,-0.13)--(1.61,-0.51)--(1.01,-0.51)--(1.01,-0.13);
    \draw[thick,fill=gray!10!white] (1.61,-0.13)--(2.4,-0.13)--(2.4,-0.57)--(1.61,-0.57)--(1.61,-0.13);
    \draw[thick,fill=gray!10!white] (2.4,-0.13)--(3.2,-0.13)--(3.2,-0.45)--(2.4,-0.45)--(2.4,-0.13);
    
    \draw[thick,fill=gray!10!white] (4.89,0.47)--(4.29,0.47) to[out=90,in=0] (3.89,0.87)--(2.89,0.87)to[out=180,in=-90](2.09,1.67)to[out=90,in=180](2.89,2.47)--(3.89,2.47) to[out=0,in=-90](4.29,2.87)--(4.89,2.87)--(4.89,0.47);

    \draw[thick,fill=gray!10!white] (4.89,-0.77)--(4.29,-0.77) to[out=-90,in=0] (3.89,-1.17)--(2.89,-1.17)to[out=180,in=90](2.09,-1.97)to[out=-90,in=180](2.89,-2.77)--(3.89,-2.77) to[out=0,in=90](4.29,-3.17)--(4.89,-3.17)--(4.89,-0.77);
    
    \draw[thick,fill=gray!30!white] (0,0)--(0,0.69)--(-0.13,0.69)--(-0.13,-0.24)--(0,-0.24)--(0,-0.13)--(5.22,-0.13)--(5.22,-3.58)--(0,-3.58)--(0,-1.18)--(-0.13,-1.18)--(-0.13,-3.71)--(5.35,-3.71)--(5.35,3.43)--(-0.13,3.43)--(-0.13,1.58)--(0,1.58)--(0,3.3)--(5.22,3.3)--(5.22,0)--(0,0);

    \draw[thick,fill=black,rotate around={-105:(0.002,0.69)}] (-0.002,0.69)--(0.002,0.69)--(0.002,1.58)--(-0.002,1.58)--(-0.002,0.69);
    \draw[thick,fill=black,rotate around={85:(0.002,-0.24)}] (-0.002,-0.24)--(0.002,-0.24)--(0.002,-1.18)--(-0.002,-1.18)--(-0.002,-0.24);
    
    \draw[fill=black] (4.69,1.67)--++(0,-0.12)--++(-0.15,0)--++(0,0.24)--++(0.15,0)--++(0,-0.12)
    ++(0,0.0075)--++(0.06,0)--++(0,-0.015)--++(-0.06,0)++(0,0.0075)
    ++(0,0.1075)--++(0.06,0)--++(0,-0.015)--++(-0.06,0)++(0,0.0075)
    ++(0,-0.1925)--++(0.06,0)--++(0,-0.015)--++(-0.06,0)++(0,0.0075);

    \draw[fill=black] (-0.25,1.82)--++(0,-0.12)--++(-0.15,0)--++(0,0.24)--++(0.15,0)--++(0,-0.12)
    ++(0,0.0075)--++(0.06,0)--++(0,-0.015)--++(-0.06,0)++(0,0.0075)
    ++(0,0.1075)--++(0.06,0)--++(0,-0.015)--++(-0.06,0)++(0,0.0075)
    ++(0,-0.1925)--++(0.06,0)--++(0,-0.015)--++(-0.06,0)++(0,0.0075);

    \draw[fill=black,rotate around={-8:(4.010,1.5)}] (4.315,1.5)--++(-0.61,0)--++(0,0.02)--++(0.09,0)--++(0,0.025)--++(0.195,0)--++(0,0.06)to[out=90,in=180]++(0.02,0.02)to[out=0,in=90]++(0.02,-0.02)--++(0,-0.06)--++(0.195,0)--++(0,-0.025)--++(0.09,0)--++(0,-0.02);
    \draw[fill=black,rotate around={8:(3.4,1.5)}] (3.705,1.5)--++(-0.61,0)--++(0,0.02)--++(0.09,0)--++(0,0.025)--++(0.195,0)--++(0,0.06)to[out=90,in=180]++(0.02,0.02)to[out=0,in=90]++(0.02,-0.02)--++(0,-0.06)--++(0.195,0)--++(0,-0.025)--++(0.09,0)--++(0,-0.02);

    \draw[fill=black,rotate around={8:(3.810,1.87)}] (4.115,1.87)--++(-0.61,0)--++(0,-0.02)--++(0.09,0)--++(0,-0.025)--++(0.195,0)--++(0,-0.06)to[out=-90,in=180]++(0.02,-0.02)to[out=0,in=-90]++(0.02,0.02)--++(0,0.06)--++(0.195,0)--++(0,0.025)--++(0.09,0)--++(0,0.02);
    \draw[fill=black,rotate around={-8:(3.2,1.87)}] (3.505,1.87)--++(-0.61,0)--++(0,-0.02)--++(0.09,0)--++(0,-0.025)--++(0.195,0)--++(0,-0.06)to[out=-90,in=180]++(0.02,-0.02)to[out=0,in=-90]++(0.02,0.02)--++(0,0.06)--++(0.195,0)--++(0,0.025)--++(0.09,0)--++(0,0.02);

    \draw[fill=black,rotate around={-8:(3.985,-2.155)}] (4.29,-2.155)--++(-0.61,0)--++(0,0.02)--++(0.09,0)--++(0,0.025)--++(0.195,0)--++(0,0.06)to[out=90,in=180]++(0.02,0.02)to[out=0,in=90]++(0.02,-0.02)--++(0,-0.06)--++(0.195,0)--++(0,-0.025)--++(0.09,0)--++(0,-0.02);
    \draw[fill=black,rotate around={8:(3.375,-2.155)}] (3.68,-2.155)--++(-0.61,0)--++(0,0.02)--++(0.09,0)--++(0,0.025)--++(0.195,0)--++(0,0.06)to[out=90,in=180]++(0.02,0.02)to[out=0,in=90]++(0.02,-0.02)--++(0,-0.06)--++(0.195,0)--++(0,-0.025)--++(0.09,0)--++(0,-0.02);

    \draw[fill=black,rotate around={8:(3.985,-1.785)}] (4.29,-1.785)--++(-0.61,0)--++(0,-0.02)--++(0.09,0)--++(0,-0.025)--++(0.195,0)--++(0,-0.06)to[out=-90,in=180]++(0.02,-0.02)to[out=0,in=-90]++(0.02,0.02)--++(0,0.06)--++(0.195,0)--++(0,0.025)--++(0.09,0)--++(0,0.02);
    \draw[fill=black,rotate around={-8:(3.375,-1.785)}] (3.68,-1.785)--++(-0.61,0)--++(0,-0.02)--++(0.09,0)--++(0,-0.025)--++(0.195,0)--++(0,-0.06)to[out=-90,in=180]++(0.02,-0.02)to[out=0,in=-90]++(0.02,0.02)--++(0,0.06)--++(0.195,0)--++(0,0.025)--++(0.09,0)--++(0,0.02);

    \draw[thick,mycolor3,->] (-3,3.43)--(-3,1.165);
    \draw[thick,mycolor3,<->] (-2.97,1.135)--(0,1.135)node[above]{\footnotesize A}--(0.9,1.135)--(2.6,2.835)--(3.9,2.835)node[above]{\footnotesize B};
    \draw[thick,mycolor3,->] (-3.03,1.105)--(-3.03,-3.71);
    \draw[thick,mycolor3,<->] (-2.97,-3.71)--(-2.97,-0.74);
    \draw[thick,mycolor3,<->] (-2.94,-0.71)--(0,-0.71)node[above]{\footnotesize C}--(0.3,-0.71)--(1.3,-1.71)node[right]{\footnotesize D};
    \draw[thick,mycolor3,<->] (1.3,-1.74)--(1.3,-3.2)--(3.9,-3.2)node[above]{\footnotesize E};
    
    \draw[thick,<->] (5.22,3.63)--(0,3.63) node[above,midway] {\footnotesize \SI{5.22}{m}};
    \draw[thick,<->] (5.55,3.3)--(5.55,0) node[right,midway] {\rotatebox{90}{\footnotesize \SI{3.3}{m}}};
    \draw[thick,<->] (5.55,-0.13)--(5.55,-3.58) node[right,midway] {\rotatebox{90}{\footnotesize \SI{3.45}{m}}};
\end{tikzpicture}
		\caption{Experiments are conducted in our offices. A person moves according to the path, starting from the top, as indicated in purple. When reaching locations B and D for the first time, the person stays at these locations for a few seconds without moving.}
		\label{fig:floorplan}
	\end{figure}
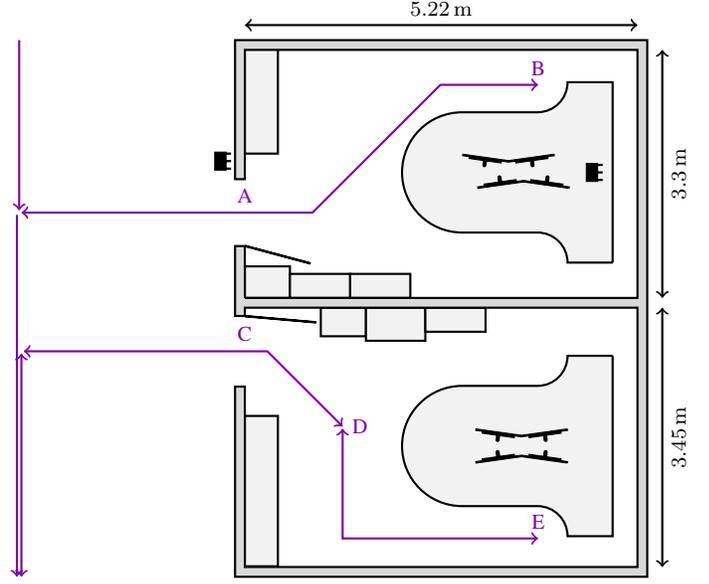%
	\begin{table}[]
		\centering
		\caption{The values of the threshold $\eta$ employed for the different feature types and algorithms, rounded.}
		\subfloat[$B=1$ bundle]{\begin{tabular}{|l|c|c|c|c|c|c|}
				\hline Prediction & \multicolumn{2}{@{}c@{}|}{Moving Avg.} & \multicolumn{2}{@{}c@{}|}{Autoregression} & \multicolumn{2}{@{}c@{}|}{Kalman}\\
				\hline $h(\bm{y}(t)|\bm{\hat{y}}_0(t))$ & \eqref{eq:h_2} & \eqref{eq:h_4} & \eqref{eq:h_2} & \eqref{eq:h_4} & \eqref{eq:h_2} & \eqref{eq:h_4} \\
				\hline f-avg. CSI std.       & 4.17 & 2.04 & 4.23 & 1.99 & 0.51 & 0.72\\
				\hline f-avg. CSI var.       & 4.21 & 2.07 & 3.84 & 1.99 & 2.22 & 1.47\\
				\hline CSI std. vec.         & 2.94 & 1.25 & 2.79 & 1.23 & 0.76 & 0.66\\
				\hline CSI var. vec.         & 2.67 & 1.22 & 2.59 & 1.19 & 4.52 & 1.32\\
				\hline t-avg. CSI ampl. vec. & 2.93 & 1.28 & 3.06 & 1.24 & 0.24 & 0.38\\
				\hline RSSI std.             & 0.27 & 0.25 & 0.34 & 0.41 & 0.06 & 0.24\\
				\hline RSSI var.             & 0.09 & 0.02 & 0.10 & 0.11 & 0.13 & 0.25\\
				\hline t-avg. RSSI           & 6.09 & 1.62 & 4.37 & 1.51 & 0.02 & 0.13\\
				\hline
		\end{tabular}}\\\vspace{0.3cm}
		
		\subfloat[$B=10$ bundles]{\begin{tabular}{|l|c|c|c|c|c|c|}
				\hline Prediction & \multicolumn{2}{@{}c@{}|}{Moving Avg.} & \multicolumn{2}{@{}c@{}|}{Autoregression} & \multicolumn{2}{@{}c@{}|}{Kalman}\\
				\hline $h(\bm{y}(t)|\bm{\hat{y}}_0(t))$ & \eqref{eq:h_2} & \eqref{eq:h_4} & \eqref{eq:h_2} & \eqref{eq:h_4} & \eqref{eq:h_2} & \eqref{eq:h_4} \\
				\hline f-avg. CSI std.       & 6.32 & 2.54 & 4.88 & 2.18 & 0.03 & 0.17\\
				\hline f-avg. CSI var.       & 6.45 & 2.43 & 4.91 & 2.12 & 0.05 & 0.23\\
				\hline CSI std. vec.         & 3.77 & 1.57 & 3.38 & 1.50 & 0.03 & 0.17\\
				\hline CSI var. vec.         & 3.80 & 1.58 & 3.38 & 1.51 & 0.07 & 0.23\\
				\hline t-avg. CSI ampl. vec. & 3.86 & 1.66 & 2.86 & 1.39 & 0.02 & 0.13\\
				\hline RSSI std.             & 9.87 & 3.37 & 6.45 & 2.75 & 0.25 & 0.53\\
				\hline RSSI var.             & 10.64 & 3.49 & 7.36 & 2.92 & 0.54 & 0.80\\
				\hline t-avg. RSSI           & 4.26 & 1.97 & 4.29 & 2.05 & 0.18 & 0.45\\
				\hline
		\end{tabular}}
		\label{tab:eta}
	\end{table}%
	The two wireless routers are set up in an office environment as shown in \figurename~\ref{fig:floorplan}; the distance between the two routers is approximately \SI{4.5}{m}. Before the experiment is started, it is ensured that no humans are present that could interfere with the measurements, and $\eta$ is set to infinity. After granting the algorithm $500$ feature measurements for convergence, a set of $500$ feature measurements without motion is recorded for initialization. The threshold value $\eta$ is chosen from the set of initial measurements as the 95-th percentile of all values of $h(\bm{y}(t)|\bm{\hat{y}}_0(t))$ recorded plus a tolerance margin of 0.1 times the difference between the 95-th percentile and the 5-th percentile. The values selected for the different combinations of feature vector and algorithm are shown in Table~\ref{tab:eta}. A person enters the area as shown with the arrow from the top in the map, and goes into the first room via position A towards position B, where the person remains without moving for approximately \SI{15}{s}. Afterwards, the person leaves the room again and leaves the area. After approximately \SI{30}{s}, the person comes back and enters the second room via position C before going to position D, where the person again remains for approximately \SI{15}{s} without moving. The person goes to position E, before going back and leaving the area again. Two cameras in the different rooms are used as a reference.
	
	\begin{figure}
		\centering
		\input{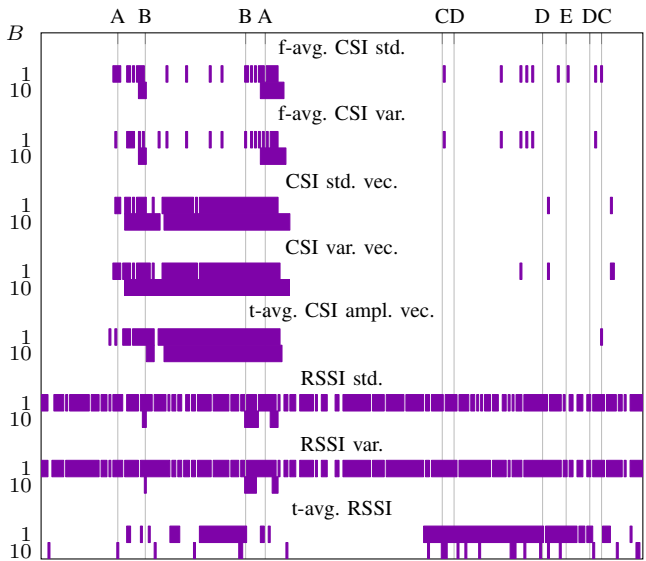}
		\caption{The time instances at which motion is detected with the moving average algorithm, hypothesis test \eqref{eq:h_2} and the different feature types.}
		\label{fig:motion_detected_MA}
	\end{figure}
	\begin{figure}
		\centering
		\input{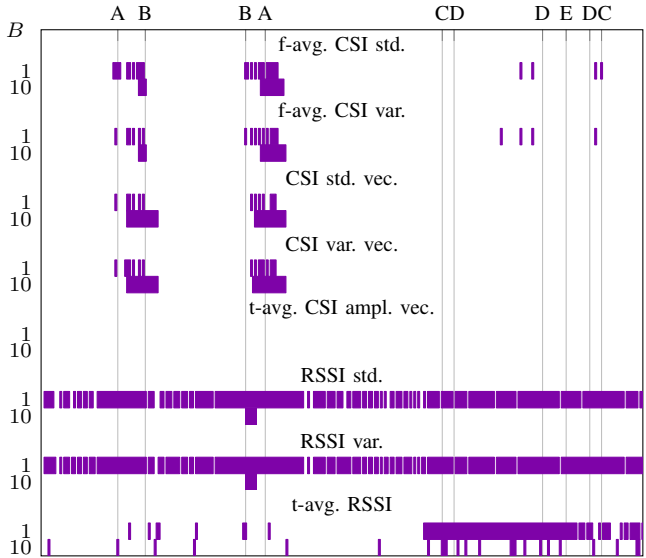}
		\caption{The time instances at which motion is detected with the moving average algorithm, hypothesis test \eqref{eq:h_4} and the different feature types.}
		\label{fig:motion_detected_MAp}
	\end{figure}
	
	\subsection{Numerical Results}
	
\begin{figure}
		\centering
		\input{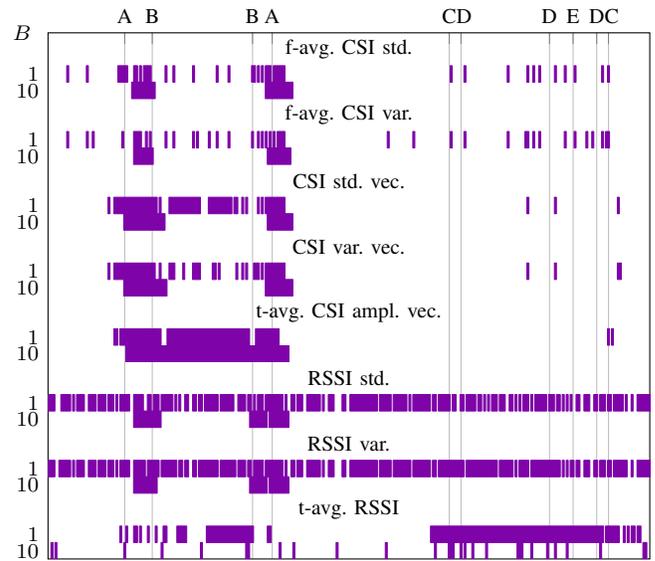}
		\caption{The time instances at which motion is detected with the autoregressive algorithm, hypothesis test \eqref{eq:h_2} and the different feature types.}
		\label{fig:motion_detected_AR}
	\end{figure}

\begin{figure}
		\centering
		\input{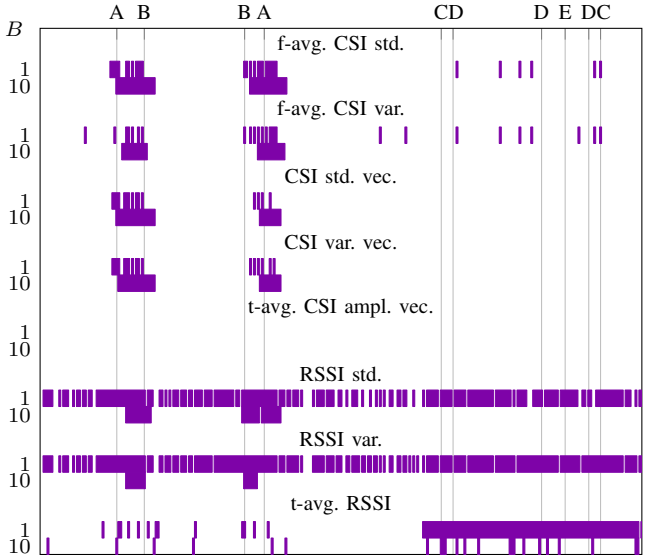}
		\caption{The time instances at which motion is detected with the autoregressive algorithm, hypothesis test \eqref{eq:h_4} and the different feature types.}
		\label{fig:motion_detected_ARp}
	\end{figure}

	The time intervals during which motion is detected with the different algorithms and feature types are shown in Figures~\ref{fig:motion_detected_MA}-\ref{fig:motion_detected_Kalmanp} in purple color. For the case of moving average prediction and omnidirectional anomaly detection, the results are shown in \figurename~\ref{fig:motion_detected_MA}. Thereby, the frequency-average CSI standard deviation and variance show the best results and detect the motion in the room closer to the devices well, but also allow recognizing minor parts of the motion related to the second room. The results of the RSSI standard deviation and variance are weaker but otherwise similar when $B=10$, but do not allow the detection of motion related to the second room. The vectors calculated from the CSI data lead to the motion being detected similarly, but also lead to false positives regarding to the pause at position B. When considering positive anomalies only as done in \figurename~\ref{fig:motion_detected_MAp}, the motion detection with all features based on standard deviation and variance of the CSI is enhanced, leading to a good recognition of motion in the first room, including a correct classification of the pause at position B. When comparing the cases of $B=1$ and $B=10$, the results of $B=10$ are more smooth, but have a slightly higher delay and almost all features do not allow the detection of motion related to the second room. Meanwhile, when only positive anomalies are considered, the motion detection based on the time-average CSI amplitude vector fails. The reason for this is that motion can lead to positive and negative anomalies in the CSI amplitude, but negative anomalies that are not tracked with hypothesis test \eqref{eq:h_4} occur more often.

	
	The results of the autoregressive prediction and omnidirectional anomaly detection in \figurename~\ref{fig:motion_detected_AR} are similar to those of the moving average prediction with the same hypothesis test design. The main differences are that the RSSI-based features show an enhanced motion detection when $B=10$ and the usage of the CSI standard deviation vector and variance vector lead to less false positives. When only measuring positive anomalies as done in \figurename~\ref{fig:motion_detected_ARp}, the time at which the person remains at position B is again almost completely correctly classified with the features based on standard deviation and variance. 
	
	\begin{figure}
		\centering
		\definecolor{mycolor1}{rgb}{0.97223,0.58172,0.25408}%
\definecolor{mycolor2}{rgb}{0.79638,0.27798,0.47132}%
\definecolor{mycolor3}{rgb}{0.49343,0.01152,0.65803}%
\definecolor{mycolor4}{rgb}{0.05038,0.02980,0.52797}%
\begin{tikzpicture}
\begin{axis}[%
width=8cm,
height=7cm,
at={(2cm,-3.75cm)},
scale only axis,
xmin=-306,
xmax=0,
ymin=-32,
ymax=0,
xtick = {-267,-253,-202,-192,-102,-96,-51,-39,-27,-21},
ytick = {0,-2.5,-3.5,-6.5,-7.5,-10.5,-11.5,-14.5,-15.5,-18.5,-19.5,-22.5,-23.5,-26.5,-27.5,-30.5,-31.5},
yticklabels = {$B\hspace{0.1cm}$,$1$,$10$,$1$,$10$,$1$,$10$,$1$,$10$,$1$,$10$,$1$,$10$,$1$,$10$,$1$,$10$},
ytick style={draw=none},
%ylabel={$B$},
%ytick = {1},%{-0.5, -1.5, -2.5, -3.5, -4.5, -5.5, -6.5, -7.5},
%yticklabels = {f-avg. CSI std., f-avg. CSI var., CSI std. vec., CSI var. vec., t-avg. CSI ampl. vec., RSSI std., RSSI var., t-avg. RSSI},
xticklabels = {A\vphantom{C},B\vphantom{C},B\vphantom{C},A\vphantom{C},C\hspace{0.08cm},\hspace{0.08cm}D\vphantom{C},D\vphantom{C},E\vphantom{C},D\vphantom{C}\hspace{0.09cm},\hspace{0.09cm}C},
ticklabel style = {font=\footnotesize},
xmajorgrids,
xtick pos=upper,xticklabel pos=upper
]
\node at ($(current axis.north)!1/32!(current axis.south)$) {\footnotesize f-avg. CSI std.};
\node at ($(current axis.north)!5/32!(current axis.south)$) {\footnotesize f-avg. CSI var.};
\node at ($(current axis.north)!9/32!(current axis.south)$) {\footnotesize CSI std. vec.};
\node at ($(current axis.north)!13/32!(current axis.south)$) {\footnotesize CSI var. vec.};
\node at ($(current axis.north)!17/32!(current axis.south)$) {\footnotesize t-avg. CSI ampl. vec.};
\node at ($(current axis.north)!21/32!(current axis.south)$) {\footnotesize RSSI std.};
\node at ($(current axis.north)!25/32!(current axis.south)$) {\footnotesize RSSI var.};
\node at ($(current axis.north)!29/32!(current axis.south)$) {\footnotesize t-avg. RSSI};

\addplot[area legend, draw=none, table/row sep=crcr, patch, patch type=rectangle, fill=mycolor3, faceted color=mycolor3, patch table={%
0	1	2	3\\
4	5	6	7\\
8	9	10	11\\
12	13	14	15\\
16	17	18	19\\
20	21	22	23\\
24	25	26	27\\
28	29	30	31\\
32	33	34	35\\
}]
table[row sep=crcr] {%
x	y\\
-285.5	-2\\
-284.5	-2\\
-284.5	-3\\
-285.5	-3\\
-260.5	-2\\
-258.5	-2\\
-258.5	-3\\
-260.5	-3\\
-213.5	-2\\
-212.5	-2\\
-212.5	-3\\
-213.5	-3\\
-197.5	-2\\
-196.5	-2\\
-196.5	-3\\
-197.5	-3\\
-195.5	-2\\
-194.5	-2\\
-194.5	-3\\
-195.5	-3\\
-193.5	-2\\
-191.5	-2\\
-191.5	-3\\
-193.5	-3\\
-189.5	-2\\
-188.5	-2\\
-188.5	-3\\
-189.5	-3\\
-145.5	-2\\
-144.5	-2\\
-144.5	-3\\
-145.5	-3\\
-56.5	-2\\
-55.5	-2\\
-55.5	-3\\
-56.5	-3\\
};

\addplot[area legend, draw=none, table/row sep=crcr, patch, patch type=rectangle, fill=mycolor3, faceted color=mycolor3, patch table={%
0	1	2	3\\
4	5	6	7\\
8	9	10	11\\
12	13	14	15\\
}]
table[row sep=crcr] {%
x	y\\
-271.5	-3\\
-254.5	-3\\
-254.5	-4\\
-271.5	-4\\
-222.5	-3\\
-221.5	-3\\
-221.5	-4\\
-222.5	-4\\
-204.5	-3\\
-185.5	-3\\
-185.5	-4\\
-204.5	-4\\
-24.5	-3\\
-23.5	-3\\
-23.5	-4\\
-24.5	-4\\
};

\addplot[area legend, draw=none, table/row sep=crcr, patch, patch type=rectangle, fill=mycolor3, faceted color=mycolor3, patch table={%
0	1	2	3\\
4	5	6	7\\
8	9	10	11\\
12	13	14	15\\
16	17	18	19\\
}]
table[row sep=crcr] {%
x	y\\
-260.5	-6\\
-251.5	-6\\
-251.5	-7\\
-260.5	-7\\
-213.5	-6\\
-210.5	-6\\
-210.5	-7\\
-213.5	-7\\
-197.5	-6\\
-187.5	-6\\
-187.5	-7\\
-197.5	-7\\
-145.5	-6\\
-142.5	-6\\
-142.5	-7\\
-145.5	-7\\
-56.5	-6\\
-54.5	-6\\
-54.5	-7\\
-56.5	-7\\
};

\addplot[area legend, draw=none, table/row sep=crcr, patch, patch type=rectangle, fill=mycolor3, faceted color=mycolor3, patch table={%
0	1	2	3\\
4	5	6	7\\
8	9	10	11\\
12	13	14	15\\
16	17	18	19\\
20	21	22	23\\
24	25	26	27\\
28	29	30	31\\
}]
table[row sep=crcr] {%
x	y\\
-271.5	-7\\
-265.5	-7\\
-265.5	-8\\
-271.5	-8\\
-262.5	-7\\
-255.5	-7\\
-255.5	-8\\
-262.5	-8\\
-222.5	-7\\
-221.5	-7\\
-221.5	-8\\
-222.5	-8\\
-210.5	-7\\
-208.5	-7\\
-208.5	-8\\
-210.5	-8\\
-202.5	-7\\
-196.5	-7\\
-196.5	-8\\
-202.5	-8\\
-195.5	-7\\
-192.5	-7\\
-192.5	-8\\
-195.5	-8\\
-188.5	-7\\
-187.5	-7\\
-187.5	-8\\
-188.5	-8\\
-183.5	-7\\
-174.5	-7\\
-174.5	-8\\
-183.5	-8\\
};

\addplot[area legend, draw=none, table/row sep=crcr, patch, patch type=rectangle, fill=mycolor3, faceted color=mycolor3, patch table={%
0	1	2	3\\
4	5	6	7\\
8	9	10	11\\
12	13	14	15\\
16	17	18	19\\
20	21	22	23\\
24	25	26	27\\
28	29	30	31\\
32	33	34	35\\
}]
table[row sep=crcr] {%
x	y\\
-275.5	-10\\
-274.5	-10\\
-274.5	-11\\
-275.5	-11\\
-271.5	-10\\
-258.5	-10\\
-258.5	-11\\
-271.5	-11\\
-256.5	-10\\
-254.5	-10\\
-254.5	-11\\
-256.5	-11\\
-249.5	-10\\
-245.5	-10\\
-245.5	-11\\
-249.5	-11\\
-231.5	-10\\
-230.5	-10\\
-230.5	-11\\
-231.5	-11\\
-213.5	-10\\
-212.5	-10\\
-212.5	-11\\
-213.5	-11\\
-203.5	-10\\
-198.5	-10\\
-198.5	-11\\
-203.5	-11\\
-197.5	-10\\
-190.5	-10\\
-190.5	-11\\
-197.5	-11\\
-189.5	-10\\
-188.5	-10\\
-188.5	-11\\
-189.5	-11\\
};

\addplot[area legend, draw=none, table/row sep=crcr, patch, patch type=rectangle, fill=mycolor3, faceted color=mycolor3, patch table={%
0	1	2	3\\
4	5	6	7\\
8	9	10	11\\
12	13	14	15\\
}]
table[row sep=crcr] {%
x	y\\
-276.5	-11\\
-158.5	-11\\
-158.5	-12\\
-276.5	-12\\
-107.5	-11\\
-83.5	-11\\
-83.5	-12\\
-107.5	-12\\
-30.5	-11\\
-27.5	-11\\
-27.5	-12\\
-30.5	-12\\
-24.5	-11\\
0.5	-11\\
0.5	-12\\
-24.5	-12\\
};

\addplot[area legend, draw=none, table/row sep=crcr, patch, patch type=rectangle, fill=mycolor3, faceted color=mycolor3, patch table={%
0	1	2	3\\
4	5	6	7\\
8	9	10	11\\
}]
table[row sep=crcr] {%
x	y\\
-275.5	-14\\
-235.5	-14\\
-235.5	-15\\
-275.5	-15\\
-231.5	-14\\
-225.5	-14\\
-225.5	-15\\
-231.5	-15\\
-203.5	-14\\
-173.5	-14\\
-173.5	-15\\
-203.5	-15\\
};

\addplot[area legend, draw=none, table/row sep=crcr, patch, patch type=rectangle, fill=mycolor3, faceted color=mycolor3, patch table={%
0	1	2	3\\
4	5	6	7\\
8	9	10	11\\
}]
table[row sep=crcr] {%
x	y\\
-276.5	-15\\
-164.5	-15\\
-164.5	-16\\
-276.5	-16\\
-108.5	-15\\
-75.5	-15\\
-75.5	-16\\
-108.5	-16\\
-24.5	-15\\
0.5	-15\\
0.5	-16\\
-24.5	-16\\
};

\addplot[area legend, draw=none, table/row sep=crcr, patch, patch type=rectangle, fill=mycolor3, faceted color=mycolor3, patch table={%
0	1	2	3\\
4	5	6	7\\
8	9	10	11\\
12	13	14	15\\
16	17	18	19\\
20	21	22	23\\
}]
table[row sep=crcr] {%
x	y\\
-276.5	-18\\
-235.5	-18\\
-235.5	-19\\
-276.5	-19\\
-233.5	-18\\
-218.5	-18\\
-218.5	-19\\
-233.5	-19\\
-203.5	-18\\
-179.5	-18\\
-179.5	-19\\
-203.5	-19\\
-107.5	-18\\
-96.5	-18\\
-96.5	-19\\
-107.5	-19\\
-63.5	-18\\
-62.5	-18\\
-62.5	-19\\
-63.5	-19\\
-21.5	-18\\
-6.5	-18\\
-6.5	-19\\
-21.5	-19\\
};

\addplot[area legend, draw=none, table/row sep=crcr, patch, patch type=rectangle, fill=mycolor3, faceted color=mycolor3, patch table={%
0	1	2	3\\
4	5	6	7\\
}]
table[row sep=crcr] {%
x	y\\
-275.5	-19\\
-166.5	-19\\
-166.5	-20\\
-275.5	-20\\
-106.5	-19\\
-80.5	-19\\
-80.5	-20\\
-106.5	-20\\
};

\addplot[area legend, draw=none, table/row sep=crcr, patch, patch type=rectangle, fill=mycolor3, faceted color=mycolor3, patch table={%
0	1	2	3\\
4	5	6	7\\
8	9	10	11\\
12	13	14	15\\
16	17	18	19\\
20	21	22	23\\
24	25	26	27\\
28	29	30	31\\
32	33	34	35\\
36	37	38	39\\
}]
table[row sep=crcr] {%
x	y\\
-304.5	-22\\
-303.5	-22\\
-303.5	-23\\
-304.5	-23\\
-262.5	-22\\
-261.5	-22\\
-261.5	-23\\
-262.5	-23\\
-254.5	-22\\
-253.5	-22\\
-253.5	-23\\
-254.5	-23\\
-195.5	-22\\
-194.5	-22\\
-194.5	-23\\
-195.5	-23\\
-193.5	-22\\
-192.5	-22\\
-192.5	-23\\
-193.5	-23\\
-191.5	-22\\
-190.5	-22\\
-190.5	-23\\
-191.5	-23\\
-187.5	-22\\
-186.5	-22\\
-186.5	-23\\
-187.5	-23\\
-116.5	-22\\
-115.5	-22\\
-115.5	-23\\
-116.5	-23\\
-107.5	-22\\
-105.5	-22\\
-105.5	-23\\
-107.5	-23\\
-27.5	-22\\
-26.5	-22\\
-26.5	-23\\
-27.5	-23\\
};

\addplot[area legend, draw=none, fill=mycolor3]
table[row sep=crcr] {%
x	y\\
100	100\\
100	100\\
150	150\\
}--cycle;

\addplot[area legend, draw=none, table/row sep=crcr, patch, patch type=rectangle, fill=mycolor3, faceted color=mycolor3, patch table={%
0	1	2	3\\
4	5	6	7\\
8	9	10	11\\
12	13	14	15\\
16	17	18	19\\
}]
table[row sep=crcr] {%
x	y\\
-262.5	-26\\
-261.5	-26\\
-261.5	-27\\
-262.5	-27\\
-254.5	-26\\
-253.5	-26\\
-253.5	-27\\
-254.5	-27\\
-195.5	-26\\
-194.5	-26\\
-194.5	-27\\
-195.5	-27\\
-193.5	-26\\
-192.5	-26\\
-192.5	-27\\
-193.5	-27\\
-191.5	-26\\
-190.5	-26\\
-190.5	-27\\
-191.5	-27\\
};

\addplot[area legend, draw=none, fill=mycolor3]
table[row sep=crcr] {%
x	y\\
100	100\\
100	100\\
150	150\\
}--cycle;

\addplot[area legend, draw=none, table/row sep=crcr, patch, patch type=rectangle, fill=mycolor3, faceted color=mycolor3, patch table={%
0	1	2	3\\
4	5	6	7\\
8	9	10	11\\
12	13	14	15\\
16	17	18	19\\
20	21	22	23\\
24	25	26	27\\
28	29	30	31\\
32	33	34	35\\
36	37	38	39\\
40	41	42	43\\
}]
table[row sep=crcr] {%
x	y\\
-267.5	-30\\
-266.5	-30\\
-266.5	-31\\
-267.5	-31\\
-261.5	-30\\
-260.5	-30\\
-260.5	-31\\
-261.5	-31\\
-256.5	-30\\
-255.5	-30\\
-255.5	-31\\
-256.5	-31\\
-254.5	-30\\
-253.5	-30\\
-253.5	-31\\
-254.5	-31\\
-252.5	-30\\
-250.5	-30\\
-250.5	-31\\
-252.5	-31\\
-248.5	-30\\
-247.5	-30\\
-247.5	-31\\
-248.5	-31\\
-203.5	-30\\
-201.5	-30\\
-201.5	-31\\
-203.5	-31\\
-197.5	-30\\
-196.5	-30\\
-196.5	-31\\
-197.5	-31\\
-192.5	-30\\
-191.5	-30\\
-191.5	-31\\
-192.5	-31\\
-190.5	-30\\
-189.5	-30\\
-189.5	-31\\
-190.5	-31\\
-111.5	-30\\
-110.5	-30\\
-110.5	-31\\
-111.5	-31\\
};

\addplot[area legend, draw=none, table/row sep=crcr, patch, patch type=rectangle, fill=mycolor3, faceted color=mycolor3, patch table={%
0	1	2	3\\
4	5	6	7\\
8	9	10	11\\
}]
table[row sep=crcr] {%
x	y\\
-279.5	-31\\
-277.5	-31\\
-277.5	-32\\
-279.5	-32\\
-248.5	-31\\
-246.5	-31\\
-246.5	-32\\
-248.5	-32\\
-169.5	-31\\
-167.5	-31\\
-167.5	-32\\
-169.5	-32\\
};

\end{axis}
\end{tikzpicture}
		\caption{The time instances at which motion is detected with the Kalman algorithm, hypothesis test \eqref{eq:h_2} and the different feature types.}
		\label{fig:motion_detected_Kalman}
	\end{figure}
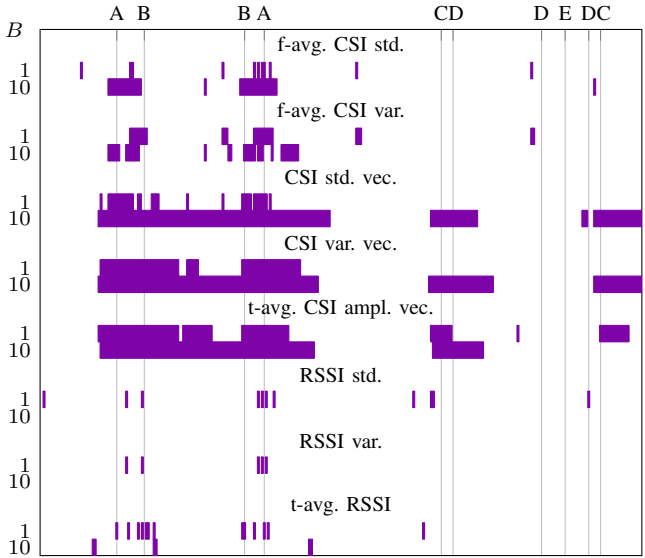
	\begin{figure}
		\centering
		\input{images/motion_detected_Kalmanp}
		\caption{The time instances at which motion is detected with the Kalman algorithm, hypothesis test \eqref{eq:h_4} and the different feature types.}
		\label{fig:motion_detected_Kalmanp}
	\end{figure}
	
	\begin{table*}[]
		\centering
		\caption{The average computing time required for the different algorithms in milliseconds for $B=1$.}
		\begin{tabular}{|l|c|c|c|c|c|c|}
			\hline Prediction & \multicolumn{2}{c|}{Moving Avg.} & \multicolumn{2}{c|}{Autoregression} & \multicolumn{2}{c|}{Kalman}\\
			\hline $h(\bm{y}(t)|\bm{\hat{y}}_0(t))$ & \eqref{eq:h_2} & \eqref{eq:h_4} & \eqref{eq:h_2} & \eqref{eq:h_4} & \eqref{eq:h_2} & \eqref{eq:h_4} \\
			\hline f-avg. CSI std.       & 0.1286 & 0.1142 & 0.0066 & 0.0055 & 0.0054 & 0.0048\\
			\hline f-avg. CSI var.       & 0.0231 & 0.0182 & 0.0059 & 0.0046 & 0.0033 & 0.0029\\
			\hline CSI std. vec.         & 0.4904 & 0.3740 & 0.0141 & 0.0116 & 0.0141 & 0.0132\\
			\hline CSI var. vec.         & 0.5274 & 0.4494 & 0.0134 & 0.0113 & 0.0132 & 0.0103\\
			\hline t-avg. CSI ampl. vec. & 0.5241 & 0.4462 & 0.0111 & 0.0095 & 0.0100 & 0.0096\\
			\hline RSSI std.             & 0.1031 & 0.0938 & 0.0054 & 0.0045 & 0.0040 & 0.0037\\
			\hline RSSI var.             & 0.0205 & 0.0163 & 0.0051 & 0.0042 & 0.0032 & 0.0027\\
			\hline t-avg. RSSI           & 0.0175 & 0.0139 & 0.0042 & 0.0036 & 0.0028 & 0.0028\\
			\hline
		\end{tabular}
		\label{tab:compTime}
	\end{table*}
	
	The results of the case where the Kalman filter is employed for feature estimation and omnidirectional anomalies are considered is shown in \figurename~\ref{fig:motion_detected_Kalman}. While a few false positives occur here, some of the CSI-based features also recognize the entering and leaving of the second room. If only positive anomalies are detected as shown in \figurename~\ref{fig:motion_detected_Kalmanp}, the CSI standard deviation and variance vectors with $B=10$ allow an even more precise detection of the entering and leaving of the second room. Parts of this is also detected with the RSSI standard deviation and variance.

	The average computing time of the different algorithms (not including the time required for the generation of the feature measurements) is shown in Tab.~\ref{tab:compTime}. All of the algorithms require less than \SI{2}{ms} per execution in our setup and are thus relatively fast. The measurements show that the autoregressive prediction is significantly faster than the moving average prediction. Moreover, the features consisting only of a scalar are faster to process, and the test design \eqref{eq:h_4} is often faster than \eqref{eq:h_2}. The fast computing times are one of the main enablers that allowed us to evaluate each data feature directly after the recording of each bundle of channel data for the results shown.
	
	\subsection{Countermeasures}
	
	To protect against sensing attacks, different measures can be used such that the feature measurements at the adversary become more similar in the cases that motion is present and not present. An approach for this can be to randomize the transmit signal gain such that the channel measurements contain similar effects of randomization as in the case of available motion. Even though a complete randomization is likely not possible for practical reasons, the case that the signal amplitude cannot be used anymore for motion detecting indicates a bound of such approaches. In this case, the RSSI-based features cannot be used. The CSI-based features can still be used if calculated based on normalized CSI vectors, which, however, reduces the information included. The above setup is utillized to evaluate features based on normalized CSI simultaneously and based on the same CSI measurements as the previous plots. The thresholds calculated from the initialization are shown in Tab.~\ref{tab:eta_norm}. Thereby, it is important to note that the overall variations of the features are reduced, such that the threshold values are lower. The values 0.00 occur due to rounding, all of the threshold values are non-negative. 
	\begin{table}[]
		\centering
		\caption{The values of the threshold $\eta$ employed for the different feature types based on normalized CSI and algorithms, rounded.}
		\subfloat[$B=1$ bundle]{\begin{tabular}{|l|c|c|c|c|c|c|}
				\hline Prediction & \multicolumn{2}{@{}c@{}|}{Moving Avg.} & \multicolumn{2}{@{}c@{}|}{Autoregression} & \multicolumn{2}{@{}c@{}|}{Kalman}\\
				\hline $h(\bm{y}(t)|\bm{\hat{y}}_0(t))$ & \eqref{eq:h_2} & \eqref{eq:h_4} & \eqref{eq:h_2} & \eqref{eq:h_4} & \eqref{eq:h_2} & \eqref{eq:h_4} \\
				\hline f-avg. CSI std.       & 0.00 & 0.04 & 0.00 & 0.04 & 0.00 & 0.00\\
				\hline f-avg. CSI var.       & 0.00 & 0.00 & 0.00 & 0.00 & 0.00 & 0.00\\
				\hline CSI std. vec.         & 0.00 & 0.04 & 0.00 & 0.04 & 0.00 & 0.00\\
				\hline CSI var. vec.         & 0.00 & 0.00 & 0.00 & 0.00 & 0.00 & 0.00\\
				\hline t-avg. CSI ampl. vec. & 0.01 & 0.02 & 0.01 & 0.02 & 0.00 & 0.01\\
				\hline
		\end{tabular}}\\\vspace{0.3cm}
		\subfloat[$B=10$ bundles]{\begin{tabular}{|l|c|c|c|c|c|c|}
				\hline Prediction & \multicolumn{2}{@{}c@{}|}{Moving Avg.} & \multicolumn{2}{@{}c@{}|}{Autoregression} & \multicolumn{2}{@{}c@{}|}{Kalman}\\
				\hline $h(\bm{y}(t)|\bm{\hat{y}}_0(t))$ & \eqref{eq:h_2} & \eqref{eq:h_4} & \eqref{eq:h_2} & \eqref{eq:h_4} & \eqref{eq:h_2} & \eqref{eq:h_4} \\
				\hline f-avg. CSI std.       & 0.00 & 0.01 & 0.00 & 0.01 & 0.00 & 0.00\\
				\hline f-avg. CSI var.       & 0.00 & 0.00 & 0.00 & 0.00 & 0.00 & 0.00\\
				\hline CSI std. vec.         & 0.00 & 0.01 & 0.00 & 0.01 & 0.00 & 0.00\\
				\hline CSI var. vec.         & 0.00 & 0.00 & 0.00 & 0.00 & 0.00 & 0.00\\
				\hline t-avg. CSI ampl. vec. & 0.00 & 0.00 & 0.00 & 0.00 & 0.00 & 0.01\\
				\hline
		\end{tabular}}
		\label{tab:eta_norm}
	\end{table}%
	\begin{figure}
		\centering
		\input{images/motion_detected_norm_MA}
		\caption{The time instances at which motion is detected with the moving average algorithm, hypothesis test \eqref{eq:h_2} and the different feature types when only normalized CSI is available.}
		\label{fig:motion_detected_norm_MA}
	\end{figure}%
	\begin{figure}
		\centering
		\definecolor{mycolor1}{rgb}{0.97223,0.58172,0.25408}%
\definecolor{mycolor2}{rgb}{0.79638,0.27798,0.47132}%
\definecolor{mycolor3}{rgb}{0.49343,0.01152,0.65803}%
\definecolor{mycolor4}{rgb}{0.05038,0.02980,0.52797}%
\begin{tikzpicture}
\begin{axis}[%
width=8cm,
height=4.375cm,
at={(2cm,-3.75cm)},
scale only axis,
xmin=-306,
xmax=0,
ymin=-20,
ymax=0,
xtick = {-267,-253,-202,-192,-102,-96,-51,-39,-27,-21},
ytick = {0,-2.5,-3.5,-6.5,-7.5,-10.5,-11.5,-14.5,-15.5,-18.5,-19.5},
yticklabels = {$B\hspace{0.1cm}$,$1$,$10$,$1$,$10$,$1$,$10$,$1$,$10$,$1$,$10$},
ytick style={draw=none},
%ylabel={$B$},
%ytick = {1},%{-0.5, -1.5, -2.5, -3.5, -4.5, -5.5, -6.5, -7.5},
%yticklabels = {f-avg. CSI std., f-avg. CSI var., CSI std. vec., CSI var. vec., t-avg. CSI ampl. vec., RSSI std., RSSI var., t-avg. RSSI},
xticklabels = {A\vphantom{C},B\vphantom{C},B\vphantom{C},A\vphantom{C},C\hspace{0.08cm},\hspace{0.08cm}D\vphantom{C},D\vphantom{C},E\vphantom{C},D\vphantom{C}\hspace{0.09cm},\hspace{0.09cm}C},
ticklabel style = {font=\footnotesize},
xmajorgrids,
xtick pos=upper,xticklabel pos=upper
]
\node at ($(current axis.north)!1/20!(current axis.south)$) {\footnotesize f-avg. CSI std.};
\node at ($(current axis.north)!5/20!(current axis.south)$) {\footnotesize f-avg. CSI var.};
\node at ($(current axis.north)!9/20!(current axis.south)$) {\footnotesize CSI std. vec.};
\node at ($(current axis.north)!13/20!(current axis.south)$) {\footnotesize CSI var. vec.};
\node at ($(current axis.north)!17/20!(current axis.south)$) {\footnotesize t-avg. CSI ampl. vec.};

\addplot[area legend, draw=none, table/row sep=crcr, patch, patch type=rectangle, fill=mycolor3, faceted color=mycolor3, patch table={%
0	1	2	3\\
4	5	6	7\\
8	9	10	11\\
12	13	14	15\\
16	17	18	19\\
20	21	22	23\\
24	25	26	27\\
28	29	30	31\\
32	33	34	35\\
36	37	38	39\\
40	41	42	43\\
44	45	46	47\\
}]
table[row sep=crcr] {%
x	y\\
-274.5	-2\\
-273.5	-2\\
-273.5	-3\\
-274.5	-3\\
-268.5	-2\\
-260.5	-2\\
-260.5	-3\\
-268.5	-3\\
-259.5	-2\\
-258.5	-2\\
-258.5	-3\\
-259.5	-3\\
-257.5	-2\\
-252.5	-2\\
-252.5	-3\\
-257.5	-3\\
-249.5	-2\\
-248.5	-2\\
-248.5	-3\\
-249.5	-3\\
-202.5	-2\\
-198.5	-2\\
-198.5	-3\\
-202.5	-3\\
-197.5	-2\\
-196.5	-2\\
-196.5	-3\\
-197.5	-3\\
-195.5	-2\\
-194.5	-2\\
-194.5	-3\\
-195.5	-3\\
-193.5	-2\\
-192.5	-2\\
-192.5	-3\\
-193.5	-3\\
-191.5	-2\\
-185.5	-2\\
-185.5	-3\\
-191.5	-3\\
-94.5	-2\\
-93.5	-2\\
-93.5	-3\\
-94.5	-3\\
-72.5	-2\\
-71.5	-2\\
-71.5	-3\\
-72.5	-3\\
};

\addplot[area legend, draw=none, table/row sep=crcr, patch, patch type=rectangle, fill=mycolor3, faceted color=mycolor3, patch table={%
0	1	2	3\\
4	5	6	7\\
8	9	10	11\\
}]
table[row sep=crcr] {%
x	y\\
-274.5	-3\\
-232.5	-3\\
-232.5	-4\\
-274.5	-4\\
-202.5	-3\\
-176.5	-3\\
-176.5	-4\\
-202.5	-4\\
-18.5	-3\\
-8.5	-3\\
-8.5	-4\\
-18.5	-4\\
};

\addplot[area legend, draw=none, table/row sep=crcr, patch, patch type=rectangle, fill=mycolor3, faceted color=mycolor3, patch table={%
0	1	2	3\\
4	5	6	7\\
8	9	10	11\\
12	13	14	15\\
16	17	18	19\\
20	21	22	23\\
24	25	26	27\\
28	29	30	31\\
32	33	34	35\\
36	37	38	39\\
40	41	42	43\\
44	45	46	47\\
48	49	50	51\\
52	53	54	55\\
56	57	58	59\\
60	61	62	63\\
}]
table[row sep=crcr] {%
x	y\\
-274.5	-6\\
-273.5	-6\\
-273.5	-7\\
-274.5	-7\\
-268.5	-6\\
-263.5	-6\\
-263.5	-7\\
-268.5	-7\\
-262.5	-6\\
-260.5	-6\\
-260.5	-7\\
-262.5	-7\\
-257.5	-6\\
-255.5	-6\\
-255.5	-7\\
-257.5	-7\\
-254.5	-6\\
-252.5	-6\\
-252.5	-7\\
-254.5	-7\\
-249.5	-6\\
-248.5	-6\\
-248.5	-7\\
-249.5	-7\\
-202.5	-6\\
-198.5	-6\\
-198.5	-7\\
-202.5	-7\\
-197.5	-6\\
-196.5	-6\\
-196.5	-7\\
-197.5	-7\\
-195.5	-6\\
-194.5	-6\\
-194.5	-7\\
-195.5	-7\\
-193.5	-6\\
-192.5	-6\\
-192.5	-7\\
-193.5	-7\\
-191.5	-6\\
-185.5	-6\\
-185.5	-7\\
-191.5	-7\\
-158.5	-6\\
-157.5	-6\\
-157.5	-7\\
-158.5	-7\\
-94.5	-6\\
-93.5	-6\\
-93.5	-7\\
-94.5	-7\\
-72.5	-6\\
-71.5	-6\\
-71.5	-7\\
-72.5	-7\\
-62.5	-6\\
-61.5	-6\\
-61.5	-7\\
-62.5	-7\\
-42.5	-6\\
-41.5	-6\\
-41.5	-7\\
-42.5	-7\\
};

\addplot[area legend, draw=none, table/row sep=crcr, patch, patch type=rectangle, fill=mycolor3, faceted color=mycolor3, patch table={%
0	1	2	3\\
4	5	6	7\\
8	9	10	11\\
12	13	14	15\\
16	17	18	19\\
20	21	22	23\\
}]
table[row sep=crcr] {%
x	y\\
-273.5	-7\\
-272.5	-7\\
-272.5	-8\\
-273.5	-8\\
-271.5	-7\\
-239.5	-7\\
-239.5	-8\\
-271.5	-8\\
-200.5	-7\\
-176.5	-7\\
-176.5	-8\\
-200.5	-8\\
-21.5	-7\\
-20.5	-7\\
-20.5	-8\\
-21.5	-8\\
-18.5	-7\\
-11.5	-7\\
-11.5	-8\\
-18.5	-8\\
-4.5	-7\\
-3.5	-7\\
-3.5	-8\\
-4.5	-8\\
};

\addplot[area legend, draw=none, table/row sep=crcr, patch, patch type=rectangle, fill=mycolor3, faceted color=mycolor3, patch table={%
0	1	2	3\\
4	5	6	7\\
8	9	10	11\\
12	13	14	15\\
16	17	18	19\\
20	21	22	23\\
24	25	26	27\\
28	29	30	31\\
32	33	34	35\\
36	37	38	39\\
40	41	42	43\\
44	45	46	47\\
}]
table[row sep=crcr] {%
x	y\\
-274.5	-10\\
-273.5	-10\\
-273.5	-11\\
-274.5	-11\\
-268.5	-10\\
-260.5	-10\\
-260.5	-11\\
-268.5	-11\\
-259.5	-10\\
-258.5	-10\\
-258.5	-11\\
-259.5	-11\\
-257.5	-10\\
-252.5	-10\\
-252.5	-11\\
-257.5	-11\\
-249.5	-10\\
-248.5	-10\\
-248.5	-11\\
-249.5	-11\\
-202.5	-10\\
-198.5	-10\\
-198.5	-11\\
-202.5	-11\\
-197.5	-10\\
-196.5	-10\\
-196.5	-11\\
-197.5	-11\\
-195.5	-10\\
-194.5	-10\\
-194.5	-11\\
-195.5	-11\\
-193.5	-10\\
-192.5	-10\\
-192.5	-11\\
-193.5	-11\\
-191.5	-10\\
-185.5	-10\\
-185.5	-11\\
-191.5	-11\\
-94.5	-10\\
-93.5	-10\\
-93.5	-11\\
-94.5	-11\\
-72.5	-10\\
-71.5	-10\\
-71.5	-11\\
-72.5	-11\\
};

\addplot[area legend, draw=none, table/row sep=crcr, patch, patch type=rectangle, fill=mycolor3, faceted color=mycolor3, patch table={%
0	1	2	3\\
4	5	6	7\\
8	9	10	11\\
}]
table[row sep=crcr] {%
x	y\\
-274.5	-11\\
-232.5	-11\\
-232.5	-12\\
-274.5	-12\\
-202.5	-11\\
-176.5	-11\\
-176.5	-12\\
-202.5	-12\\
-18.5	-11\\
-8.5	-11\\
-8.5	-12\\
-18.5	-12\\
};

\addplot[area legend, draw=none, table/row sep=crcr, patch, patch type=rectangle, fill=mycolor3, faceted color=mycolor3, patch table={%
0	1	2	3\\
4	5	6	7\\
8	9	10	11\\
12	13	14	15\\
16	17	18	19\\
20	21	22	23\\
24	25	26	27\\
28	29	30	31\\
32	33	34	35\\
36	37	38	39\\
40	41	42	43\\
44	45	46	47\\
48	49	50	51\\
52	53	54	55\\
56	57	58	59\\
60	61	62	63\\
}]
table[row sep=crcr] {%
x	y\\
-274.5	-14\\
-273.5	-14\\
-273.5	-15\\
-274.5	-15\\
-268.5	-14\\
-263.5	-14\\
-263.5	-15\\
-268.5	-15\\
-262.5	-14\\
-260.5	-14\\
-260.5	-15\\
-262.5	-15\\
-257.5	-14\\
-255.5	-14\\
-255.5	-15\\
-257.5	-15\\
-254.5	-14\\
-252.5	-14\\
-252.5	-15\\
-254.5	-15\\
-249.5	-14\\
-248.5	-14\\
-248.5	-15\\
-249.5	-15\\
-202.5	-14\\
-198.5	-14\\
-198.5	-15\\
-202.5	-15\\
-197.5	-14\\
-196.5	-14\\
-196.5	-15\\
-197.5	-15\\
-195.5	-14\\
-194.5	-14\\
-194.5	-15\\
-195.5	-15\\
-193.5	-14\\
-192.5	-14\\
-192.5	-15\\
-193.5	-15\\
-191.5	-14\\
-185.5	-14\\
-185.5	-15\\
-191.5	-15\\
-158.5	-14\\
-157.5	-14\\
-157.5	-15\\
-158.5	-15\\
-94.5	-14\\
-93.5	-14\\
-93.5	-15\\
-94.5	-15\\
-72.5	-14\\
-71.5	-14\\
-71.5	-15\\
-72.5	-15\\
-62.5	-14\\
-61.5	-14\\
-61.5	-15\\
-62.5	-15\\
-42.5	-14\\
-41.5	-14\\
-41.5	-15\\
-42.5	-15\\
};

\addplot[area legend, draw=none, table/row sep=crcr, patch, patch type=rectangle, fill=mycolor3, faceted color=mycolor3, patch table={%
0	1	2	3\\
4	5	6	7\\
8	9	10	11\\
12	13	14	15\\
16	17	18	19\\
20	21	22	23\\
}]
table[row sep=crcr] {%
x	y\\
-273.5	-15\\
-272.5	-15\\
-272.5	-16\\
-273.5	-16\\
-271.5	-15\\
-239.5	-15\\
-239.5	-16\\
-271.5	-16\\
-200.5	-15\\
-176.5	-15\\
-176.5	-16\\
-200.5	-16\\
-21.5	-15\\
-20.5	-15\\
-20.5	-16\\
-21.5	-16\\
-18.5	-15\\
-11.5	-15\\
-11.5	-16\\
-18.5	-16\\
-4.5	-15\\
-3.5	-15\\
-3.5	-16\\
-4.5	-16\\
};

\addplot[area legend, draw=none, table/row sep=crcr, patch, patch type=rectangle, fill=mycolor3, faceted color=mycolor3, patch table={%
0	1	2	3\\
4	5	6	7\\
8	9	10	11\\
12	13	14	15\\
16	17	18	19\\
20	21	22	23\\
24	25	26	27\\
28	29	30	31\\
32	33	34	35\\
36	37	38	39\\
40	41	42	43\\
44	45	46	47\\
48	49	50	51\\
52	53	54	55\\
56	57	58	59\\
60	61	62	63\\
64	65	66	67\\
68	69	70	71\\
72	73	74	75\\
76	77	78	79\\
80	81	82	83\\
84	85	86	87\\
88	89	90	91\\
}]
table[row sep=crcr] {%
x	y\\
-313.5	-18\\
-309.5	-18\\
-309.5	-19\\
-313.5	-19\\
-308.5	-18\\
-304.5	-18\\
-304.5	-19\\
-308.5	-19\\
-302.5	-18\\
-300.5	-18\\
-300.5	-19\\
-302.5	-19\\
-298.5	-18\\
-295.5	-18\\
-295.5	-19\\
-298.5	-19\\
-294.5	-18\\
-290.5	-18\\
-290.5	-19\\
-294.5	-19\\
-289.5	-18\\
-288.5	-18\\
-288.5	-19\\
-289.5	-19\\
-286.5	-18\\
-285.5	-18\\
-285.5	-19\\
-286.5	-19\\
-284.5	-18\\
-283.5	-18\\
-283.5	-19\\
-284.5	-19\\
-282.5	-18\\
-279.5	-18\\
-279.5	-19\\
-282.5	-19\\
-275.5	-18\\
-274.5	-18\\
-274.5	-19\\
-275.5	-19\\
-272.5	-18\\
-271.5	-18\\
-271.5	-19\\
-272.5	-19\\
-261.5	-18\\
-259.5	-18\\
-259.5	-19\\
-261.5	-19\\
-257.5	-18\\
-256.5	-18\\
-256.5	-19\\
-257.5	-19\\
-255.5	-18\\
-254.5	-18\\
-254.5	-19\\
-255.5	-19\\
-252.5	-18\\
-251.5	-18\\
-251.5	-19\\
-252.5	-19\\
-250.5	-18\\
-249.5	-18\\
-249.5	-19\\
-250.5	-19\\
-247.5	-18\\
-246.5	-18\\
-246.5	-19\\
-247.5	-19\\
-245.5	-18\\
-244.5	-18\\
-244.5	-19\\
-245.5	-19\\
-242.5	-18\\
-241.5	-18\\
-241.5	-19\\
-242.5	-19\\
-240.5	-18\\
-239.5	-18\\
-239.5	-19\\
-240.5	-19\\
-238.5	-18\\
-203.5	-18\\
-203.5	-19\\
-238.5	-19\\
-198.5	-18\\
-197.5	-18\\
-197.5	-19\\
-198.5	-19\\
-196.5	-18\\
-193.5	-18\\
-193.5	-19\\
-196.5	-19\\
};

\addplot[area legend, draw=none, table/row sep=crcr, patch, patch type=rectangle, fill=mycolor3, faceted color=mycolor3, patch table={%
0	1	2	3\\
4	5	6	7\\
}]
table[row sep=crcr] {%
x	y\\
-313.5	-19\\
-266.5	-19\\
-266.5	-20\\
-313.5	-20\\
-255.5	-19\\
-186.5	-19\\
-186.5	-20\\
-255.5	-20\\
};

\end{axis}
\end{tikzpicture}
		\caption{The time instances at which motion is detected with the moving average algorithm, hypothesis test \eqref{eq:h_4} and the different feature types when only normalized CSI is available.}
		\label{fig:motion_detected_norm_MAp}
	\end{figure}%
	\begin{figure}
		\centering
		\input{images/motion_detected_norm_AR}
		\caption{The time instances at which motion is detected with the autoregressive algorithm, hypothesis test \eqref{eq:h_2} and the different feature types when only normalized CSI is available.}
		\label{fig:motion_detected_norm_AR}
	\end{figure}%
	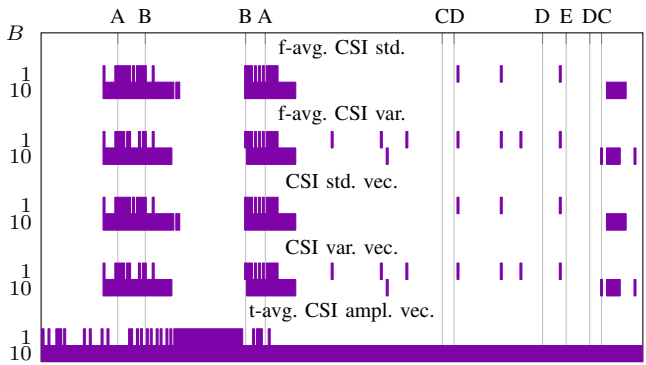
\begin{figure}
		\centering
		\definecolor{mycolor1}{rgb}{0.97223,0.58172,0.25408}%
\definecolor{mycolor2}{rgb}{0.79638,0.27798,0.47132}%
\definecolor{mycolor3}{rgb}{0.49343,0.01152,0.65803}%
\definecolor{mycolor4}{rgb}{0.05038,0.02980,0.52797}%
\begin{tikzpicture}
\begin{axis}[%
width=8cm,
height=4.375cm,
at={(2cm,-3.75cm)},
scale only axis,
xmin=-306,
xmax=0,
ymin=-20,
ymax=0,
xtick = {-267,-253,-202,-192,-102,-96,-51,-39,-27,-21},
ytick = {0,-2.5,-3.5,-6.5,-7.5,-10.5,-11.5,-14.5,-15.5,-18.5,-19.5},
yticklabels = {$B\hspace{0.1cm}$,$1$,$10$,$1$,$10$,$1$,$10$,$1$,$10$,$1$,$10$},
ytick style={draw=none},
%ylabel={$B$},
%ytick = {1},%{-0.5, -1.5, -2.5, -3.5, -4.5, -5.5, -6.5, -7.5},
%yticklabels = {f-avg. CSI std., f-avg. CSI var., CSI std. vec., CSI var. vec., t-avg. CSI ampl. vec., RSSI std., RSSI var., t-avg. RSSI},
xticklabels = {A\vphantom{C},B\vphantom{C},B\vphantom{C},A\vphantom{C},C\hspace{0.08cm},\hspace{0.08cm}D\vphantom{C},D\vphantom{C},E\vphantom{C},D\vphantom{C}\hspace{0.09cm},\hspace{0.09cm}C},
ticklabel style = {font=\footnotesize},
xmajorgrids,
xtick pos=upper,xticklabel pos=upper
]
\node at ($(current axis.north)!1/20!(current axis.south)$) {\footnotesize f-avg. CSI std.};
\node at ($(current axis.north)!5/20!(current axis.south)$) {\footnotesize f-avg. CSI var.};
\node at ($(current axis.north)!9/20!(current axis.south)$) {\footnotesize CSI std. vec.};
\node at ($(current axis.north)!13/20!(current axis.south)$) {\footnotesize CSI var. vec.};
\node at ($(current axis.north)!17/20!(current axis.south)$) {\footnotesize t-avg. CSI ampl. vec.};

\addplot[area legend, draw=none, table/row sep=crcr, patch, patch type=rectangle, fill=mycolor3, faceted color=mycolor3, patch table={%
0	1	2	3\\
4	5	6	7\\
8	9	10	11\\
12	13	14	15\\
16	17	18	19\\
20	21	22	23\\
24	25	26	27\\
28	29	30	31\\
32	33	34	35\\
36	37	38	39\\
40	41	42	43\\
44	45	46	47\\
48	49	50	51\\
}]
table[row sep=crcr] {%
x	y\\
-274.5	-2\\
-273.5	-2\\
-273.5	-3\\
-274.5	-3\\
-268.5	-2\\
-260.5	-2\\
-260.5	-3\\
-268.5	-3\\
-259.5	-2\\
-258.5	-2\\
-258.5	-3\\
-259.5	-3\\
-257.5	-2\\
-252.5	-2\\
-252.5	-3\\
-257.5	-3\\
-249.5	-2\\
-248.5	-2\\
-248.5	-3\\
-249.5	-3\\
-202.5	-2\\
-198.5	-2\\
-198.5	-3\\
-202.5	-3\\
-197.5	-2\\
-196.5	-2\\
-196.5	-3\\
-197.5	-3\\
-195.5	-2\\
-194.5	-2\\
-194.5	-3\\
-195.5	-3\\
-193.5	-2\\
-192.5	-2\\
-192.5	-3\\
-193.5	-3\\
-191.5	-2\\
-185.5	-2\\
-185.5	-3\\
-191.5	-3\\
-94.5	-2\\
-93.5	-2\\
-93.5	-3\\
-94.5	-3\\
-72.5	-2\\
-71.5	-2\\
-71.5	-3\\
-72.5	-3\\
-42.5	-2\\
-41.5	-2\\
-41.5	-3\\
-42.5	-3\\
};

\addplot[area legend, draw=none, table/row sep=crcr, patch, patch type=rectangle, fill=mycolor3, faceted color=mycolor3, patch table={%
0	1	2	3\\
4	5	6	7\\
8	9	10	11\\
12	13	14	15\\
}]
table[row sep=crcr] {%
x	y\\
-274.5	-3\\
-238.5	-3\\
-238.5	-4\\
-274.5	-4\\
-237.5	-3\\
-235.5	-3\\
-235.5	-4\\
-237.5	-4\\
-202.5	-3\\
-176.5	-3\\
-176.5	-4\\
-202.5	-4\\
-18.5	-3\\
-8.5	-3\\
-8.5	-4\\
-18.5	-4\\
};

\addplot[area legend, draw=none, table/row sep=crcr, patch, patch type=rectangle, fill=mycolor3, faceted color=mycolor3, patch table={%
0	1	2	3\\
4	5	6	7\\
8	9	10	11\\
12	13	14	15\\
16	17	18	19\\
20	21	22	23\\
24	25	26	27\\
28	29	30	31\\
32	33	34	35\\
36	37	38	39\\
40	41	42	43\\
44	45	46	47\\
48	49	50	51\\
52	53	54	55\\
56	57	58	59\\
60	61	62	63\\
64	65	66	67\\
68	69	70	71\\
}]
table[row sep=crcr] {%
x	y\\
-274.5	-6\\
-273.5	-6\\
-273.5	-7\\
-274.5	-7\\
-268.5	-6\\
-263.5	-6\\
-263.5	-7\\
-268.5	-7\\
-262.5	-6\\
-260.5	-6\\
-260.5	-7\\
-262.5	-7\\
-256.5	-6\\
-255.5	-6\\
-255.5	-7\\
-256.5	-7\\
-254.5	-6\\
-252.5	-6\\
-252.5	-7\\
-254.5	-7\\
-249.5	-6\\
-248.5	-6\\
-248.5	-7\\
-249.5	-7\\
-202.5	-6\\
-198.5	-6\\
-198.5	-7\\
-202.5	-7\\
-197.5	-6\\
-196.5	-6\\
-196.5	-7\\
-197.5	-7\\
-195.5	-6\\
-194.5	-6\\
-194.5	-7\\
-195.5	-7\\
-193.5	-6\\
-192.5	-6\\
-192.5	-7\\
-193.5	-7\\
-191.5	-6\\
-185.5	-6\\
-185.5	-7\\
-191.5	-7\\
-158.5	-6\\
-157.5	-6\\
-157.5	-7\\
-158.5	-7\\
-133.5	-6\\
-132.5	-6\\
-132.5	-7\\
-133.5	-7\\
-120.5	-6\\
-119.5	-6\\
-119.5	-7\\
-120.5	-7\\
-94.5	-6\\
-93.5	-6\\
-93.5	-7\\
-94.5	-7\\
-72.5	-6\\
-71.5	-6\\
-71.5	-7\\
-72.5	-7\\
-62.5	-6\\
-61.5	-6\\
-61.5	-7\\
-62.5	-7\\
-42.5	-6\\
-41.5	-6\\
-41.5	-7\\
-42.5	-7\\
};

\addplot[area legend, draw=none, table/row sep=crcr, patch, patch type=rectangle, fill=mycolor3, faceted color=mycolor3, patch table={%
0	1	2	3\\
4	5	6	7\\
8	9	10	11\\
12	13	14	15\\
16	17	18	19\\
20	21	22	23\\
}]
table[row sep=crcr] {%
x	y\\
-274.5	-7\\
-239.5	-7\\
-239.5	-8\\
-274.5	-8\\
-201.5	-7\\
-176.5	-7\\
-176.5	-8\\
-201.5	-8\\
-130.5	-7\\
-129.5	-7\\
-129.5	-8\\
-130.5	-8\\
-21.5	-7\\
-20.5	-7\\
-20.5	-8\\
-21.5	-8\\
-18.5	-7\\
-11.5	-7\\
-11.5	-8\\
-18.5	-8\\
-4.5	-7\\
-3.5	-7\\
-3.5	-8\\
-4.5	-8\\
};

\addplot[area legend, draw=none, table/row sep=crcr, patch, patch type=rectangle, fill=mycolor3, faceted color=mycolor3, patch table={%
0	1	2	3\\
4	5	6	7\\
8	9	10	11\\
12	13	14	15\\
16	17	18	19\\
20	21	22	23\\
24	25	26	27\\
28	29	30	31\\
32	33	34	35\\
36	37	38	39\\
40	41	42	43\\
44	45	46	47\\
48	49	50	51\\
}]
table[row sep=crcr] {%
x	y\\
-274.5	-10\\
-273.5	-10\\
-273.5	-11\\
-274.5	-11\\
-268.5	-10\\
-260.5	-10\\
-260.5	-11\\
-268.5	-11\\
-259.5	-10\\
-258.5	-10\\
-258.5	-11\\
-259.5	-11\\
-257.5	-10\\
-252.5	-10\\
-252.5	-11\\
-257.5	-11\\
-249.5	-10\\
-248.5	-10\\
-248.5	-11\\
-249.5	-11\\
-202.5	-10\\
-198.5	-10\\
-198.5	-11\\
-202.5	-11\\
-197.5	-10\\
-196.5	-10\\
-196.5	-11\\
-197.5	-11\\
-195.5	-10\\
-194.5	-10\\
-194.5	-11\\
-195.5	-11\\
-193.5	-10\\
-192.5	-10\\
-192.5	-11\\
-193.5	-11\\
-191.5	-10\\
-185.5	-10\\
-185.5	-11\\
-191.5	-11\\
-94.5	-10\\
-93.5	-10\\
-93.5	-11\\
-94.5	-11\\
-72.5	-10\\
-71.5	-10\\
-71.5	-11\\
-72.5	-11\\
-42.5	-10\\
-41.5	-10\\
-41.5	-11\\
-42.5	-11\\
};

\addplot[area legend, draw=none, table/row sep=crcr, patch, patch type=rectangle, fill=mycolor3, faceted color=mycolor3, patch table={%
0	1	2	3\\
4	5	6	7\\
8	9	10	11\\
12	13	14	15\\
}]
table[row sep=crcr] {%
x	y\\
-274.5	-11\\
-238.5	-11\\
-238.5	-12\\
-274.5	-12\\
-237.5	-11\\
-235.5	-11\\
-235.5	-12\\
-237.5	-12\\
-202.5	-11\\
-176.5	-11\\
-176.5	-12\\
-202.5	-12\\
-18.5	-11\\
-8.5	-11\\
-8.5	-12\\
-18.5	-12\\
};

\addplot[area legend, draw=none, table/row sep=crcr, patch, patch type=rectangle, fill=mycolor3, faceted color=mycolor3, patch table={%
0	1	2	3\\
4	5	6	7\\
8	9	10	11\\
12	13	14	15\\
16	17	18	19\\
20	21	22	23\\
24	25	26	27\\
28	29	30	31\\
32	33	34	35\\
36	37	38	39\\
40	41	42	43\\
44	45	46	47\\
48	49	50	51\\
52	53	54	55\\
56	57	58	59\\
60	61	62	63\\
64	65	66	67\\
68	69	70	71\\
}]
table[row sep=crcr] {%
x	y\\
-274.5	-14\\
-273.5	-14\\
-273.5	-15\\
-274.5	-15\\
-268.5	-14\\
-263.5	-14\\
-263.5	-15\\
-268.5	-15\\
-262.5	-14\\
-260.5	-14\\
-260.5	-15\\
-262.5	-15\\
-256.5	-14\\
-255.5	-14\\
-255.5	-15\\
-256.5	-15\\
-254.5	-14\\
-252.5	-14\\
-252.5	-15\\
-254.5	-15\\
-249.5	-14\\
-248.5	-14\\
-248.5	-15\\
-249.5	-15\\
-202.5	-14\\
-198.5	-14\\
-198.5	-15\\
-202.5	-15\\
-197.5	-14\\
-196.5	-14\\
-196.5	-15\\
-197.5	-15\\
-195.5	-14\\
-194.5	-14\\
-194.5	-15\\
-195.5	-15\\
-193.5	-14\\
-192.5	-14\\
-192.5	-15\\
-193.5	-15\\
-191.5	-14\\
-185.5	-14\\
-185.5	-15\\
-191.5	-15\\
-158.5	-14\\
-157.5	-14\\
-157.5	-15\\
-158.5	-15\\
-133.5	-14\\
-132.5	-14\\
-132.5	-15\\
-133.5	-15\\
-120.5	-14\\
-119.5	-14\\
-119.5	-15\\
-120.5	-15\\
-94.5	-14\\
-93.5	-14\\
-93.5	-15\\
-94.5	-15\\
-72.5	-14\\
-71.5	-14\\
-71.5	-15\\
-72.5	-15\\
-62.5	-14\\
-61.5	-14\\
-61.5	-15\\
-62.5	-15\\
-42.5	-14\\
-41.5	-14\\
-41.5	-15\\
-42.5	-15\\
};

\addplot[area legend, draw=none, table/row sep=crcr, patch, patch type=rectangle, fill=mycolor3, faceted color=mycolor3, patch table={%
0	1	2	3\\
4	5	6	7\\
8	9	10	11\\
12	13	14	15\\
16	17	18	19\\
20	21	22	23\\
}]
table[row sep=crcr] {%
x	y\\
-274.5	-15\\
-239.5	-15\\
-239.5	-16\\
-274.5	-16\\
-201.5	-15\\
-176.5	-15\\
-176.5	-16\\
-201.5	-16\\
-130.5	-15\\
-129.5	-15\\
-129.5	-16\\
-130.5	-16\\
-21.5	-15\\
-20.5	-15\\
-20.5	-16\\
-21.5	-16\\
-18.5	-15\\
-11.5	-15\\
-11.5	-16\\
-18.5	-16\\
-4.5	-15\\
-3.5	-15\\
-3.5	-16\\
-4.5	-16\\
};

\addplot[area legend, draw=none, table/row sep=crcr, patch, patch type=rectangle, fill=mycolor3, faceted color=mycolor3, patch table={%
0	1	2	3\\
4	5	6	7\\
8	9	10	11\\
12	13	14	15\\
16	17	18	19\\
20	21	22	23\\
24	25	26	27\\
28	29	30	31\\
32	33	34	35\\
36	37	38	39\\
40	41	42	43\\
44	45	46	47\\
48	49	50	51\\
52	53	54	55\\
56	57	58	59\\
60	61	62	63\\
64	65	66	67\\
68	69	70	71\\
72	73	74	75\\
76	77	78	79\\
80	81	82	83\\
84	85	86	87\\
88	89	90	91\\
}]
table[row sep=crcr] {%
x	y\\
-313.5	-18\\
-310.5	-18\\
-310.5	-19\\
-313.5	-19\\
-308.5	-18\\
-307.5	-18\\
-307.5	-19\\
-308.5	-19\\
-306.5	-18\\
-304.5	-18\\
-304.5	-19\\
-306.5	-19\\
-302.5	-18\\
-301.5	-18\\
-301.5	-19\\
-302.5	-19\\
-298.5	-18\\
-295.5	-18\\
-295.5	-19\\
-298.5	-19\\
-294.5	-18\\
-293.5	-18\\
-293.5	-19\\
-294.5	-19\\
-284.5	-18\\
-283.5	-18\\
-283.5	-19\\
-284.5	-19\\
-281.5	-18\\
-280.5	-18\\
-280.5	-19\\
-281.5	-19\\
-275.5	-18\\
-274.5	-18\\
-274.5	-19\\
-275.5	-19\\
-272.5	-18\\
-271.5	-18\\
-271.5	-19\\
-272.5	-19\\
-261.5	-18\\
-259.5	-18\\
-259.5	-19\\
-261.5	-19\\
-257.5	-18\\
-256.5	-18\\
-256.5	-19\\
-257.5	-19\\
-255.5	-18\\
-254.5	-18\\
-254.5	-19\\
-255.5	-19\\
-252.5	-18\\
-251.5	-18\\
-251.5	-19\\
-252.5	-19\\
-250.5	-18\\
-249.5	-18\\
-249.5	-19\\
-250.5	-19\\
-247.5	-18\\
-246.5	-18\\
-246.5	-19\\
-247.5	-19\\
-245.5	-18\\
-244.5	-18\\
-244.5	-19\\
-245.5	-19\\
-242.5	-18\\
-241.5	-18\\
-241.5	-19\\
-242.5	-19\\
-240.5	-18\\
-239.5	-18\\
-239.5	-19\\
-240.5	-19\\
-238.5	-18\\
-203.5	-18\\
-203.5	-19\\
-238.5	-19\\
-198.5	-18\\
-197.5	-18\\
-197.5	-19\\
-198.5	-19\\
-196.5	-18\\
-193.5	-18\\
-193.5	-19\\
-196.5	-19\\
-190.5	-18\\
-189.5	-18\\
-189.5	-19\\
-190.5	-19\\
};

\addplot[area legend, draw=none, fill=mycolor3]
table[row sep=crcr] {%
x	y\\
-313.5	-19\\
0.5	-19\\
0.5	-20\\
-313.5	-20\\
}--cycle;

\end{axis}
\end{tikzpicture}
		\caption{The time instances at which motion is detected with the autoregressive algorithm, hypothesis test \eqref{eq:h_4} and the different feature types when only normalized CSI is available.}
		\label{fig:motion_detected_norm_ARp}
	\end{figure}%
	In Figures~\ref{fig:motion_detected_norm_MA}-\ref{fig:motion_detected_norm_Kalmanp}, the obtained motion detection performance is shown. The results show that the motion detection performance with some of the features, such as the time-average CSI amplitude vector, are reduced, indicating the possible effectiveness of transmit power randomization. However, the features related to the standard deviation and variance show together with hypothesis test \eqref{eq:h_4} and the moving average or autoregressive filter a very good motion detection. The according measurement sequences even show some recognition of motion related to the second room. 
	
	This indicates that once the adversary has access to continuous measurements of CSI data, a frequency-dependent randomization might be required to obscure motion. Moreover, it needs to be handled that likely only a small amount of randomization is possible as the power needs to be balanced between targets of a reliable and energy-efficient communication, such that parts of the information on the channel gain might remain available. Therefore, attempts of randomization in other dimensions can be added, such as in the location through changes between multiple access points \cite{10.1145/3395351.3399368} or in the channel through randomization via reconfigurable intelligent surfaces \cite{Staat2021IRShieldAC}.

	\begin{figure}
		\centering
		\input{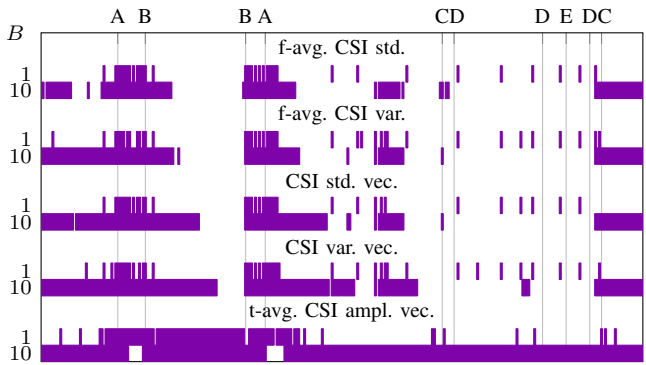}
		\caption{The time instances at which motion is detected with the Kalman algorithm, hypothesis test \eqref{eq:h_2} and the different feature types when only normalized CSI is available.}
		\label{fig:motion_detected_norm_Kalman}
	\end{figure}
	\begin{figure}
		\centering
		\input{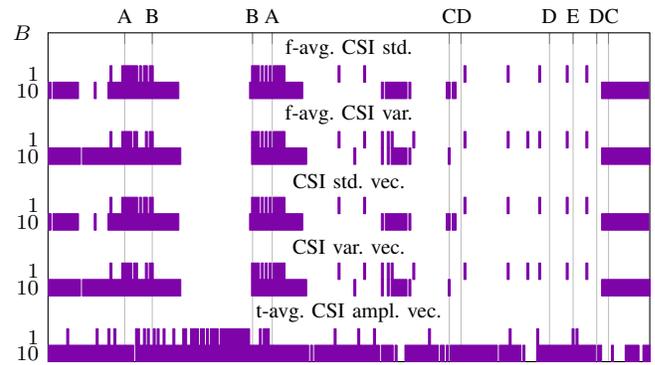}
		\caption{The time instances at which motion is detected with the Kalman algorithm, hypothesis test \eqref{eq:h_4} and the different feature types when only normalized CSI is available.}
		\label{fig:motion_detected_norm_Kalmanp}
	\end{figure}
	
	\section{Conclusion}
	
	Anomaly detection algorithms for physical layer security require real-time processing that is robust to slow-fading changes in the device's environment. The case study confirms that the proposed algorithms based on moving average predictions, autoregressive predictions and Kalman filters can be executed timely, which makes them suitable for such applications. The algorithms can be utilized to evaluate different features, such as in the case study on sensing attacks related to the channel gain or variability within the channel (such as the short-term standard variation or variance). As motion leads to increases in the variability, variability-based features show an improved detection within a hypothesis test design that only detects positive anomalies. For features related to the channel gain, an omnidirectional anomaly detection shows advantages. While a randomization of the transmit signal gain can protect users against attacks based on some of the features and algorithm, a frequency-dependent randomization might be required to protect against the utillization of all features and algorithms.

	\bibliographystyle{IEEEtran}
	\bibliography{main}
	
\end{document}